\documentclass[preprint,superscriptaddress]{revtex4-1}
\usepackage{amssymb}
\usepackage[utf8]{inputenc}
\usepackage{graphicx}
\usepackage{bm,color,subfigure,amsmath,hyperref}

\begin{document}

\title{Breaking of Galilean invariance in the hydrodynamic formulation of ferromagnetic thin films}

\author{Ezio~Iacocca}
\email[E-mail: ]{ezio.iacocca@colorado.edu}
\thanks{Videos of micromagnetic simulations \href{http://www.colorado.edu/amath/dispersive-hydrodynamics-lab/videos}{here}.}
\affiliation{Department of Applied Mathematics, University of Colorado, Boulder, Colorado 80309-0526, USA}
\affiliation{Department of Physics, Division for Theoretical Physics, Chalmers University of Technology, 412 96, Gothenburg, Sweden}

\author{T. J.~Silva}
\affiliation{National Institute of Standards and Technology, Boulder, Colorado 80305-3328, USA}
\thanks{Contribution of the National Institute of Standards and Technology; not subject to copyright in the United States.}

\author{Mark~A.~Hoefer}
\affiliation{Department of Applied Mathematics, University of Colorado, Boulder, Colorado 80309-0526, USA}

\begin{abstract}
  Microwave magnetodynamics in ferromagnets are often studied in the small-amplitude or weakly nonlinear regime
    corresponding to modulations of a well-defined
  magnetic state. However, strongly nonlinear regimes, where the
  aforementioned approximations are not applicable,
  have become experimentally accessible. By
  re-interpreting the governing Landau-Lifshitz equation of motion, we
  derive an exact set of equations of dispersive hydrodynamic form
  that are amenable to analytical study even when full nonlinearity
  and exchange dispersion are included. The resulting equations are shown to, in general, break Galilean invariance. A 
  magnetic Mach number is obtained as a function of static and moving
  reference frames. The simplest class of solutions are termed uniform
  hydrodynamic states (UHSs), which exhibit fluid-like
  behavior including laminar flow at subsonic speeds and the formation
  of a Mach cone and wave-fronts at supersonic speeds. A regime of
  modulational instability is also possible, where the UHS is
  violently unstable. The hydrodynamic interpretation opens up novel
  possibilities in magnetic research.
\end{abstract}

\maketitle

Magnetodynamics in thin film ferromagnets have been studied
for many decades. Advances in nanofabrication and the advent
of spin transfer~\cite{Berger1996,Slonczewski1996} and spin-orbit torques~\cite{Hals2013b} have opened a new frontier of experimentally accessible nonlinear
physics~\cite{Hoefer2010,Mohseni2013,Macia2014,Chung2016,Mistral2008}.
Large-amplitude excitations negate the use of typical linear or weakly
nonlinear analyses~\cite{Slavin2009,Lvov1994,Bertotti2008}, necessitating instead either
micromagnetic simulations~\cite{Vansteenkiste2014} or analytical approaches suited to strongly
nonlinear dynamics.  Therefore, an interpretation of the Landau-Lifshitz (LL) equation that includes full nonlinearity, yet is
amenable to analytical study, would be insightful.

A hydrodynamic interpretation was
proposed by Halperin and Hohenberg~\cite{Halperin1969} to describe spin waves in
anisotropic ferro- and antiferromagnets. Recently, theoretical
studies of thin film ferromagnets with planar
anisotropy have identified a relationship to superfluid-like
hydrodynamic
equations~\cite{Konig2001,Sonin2010,Takei2014,Chen2014,Skarsvaag2015,Flebus2016} supporting large-amplitude modes beyond strongly nonlinear spin wave and macrospin modes~\cite{Lvov1994,Bertotti2008}. However,
these studies are limited to the long-wavelength, low-frequency
regime where linear and weakly-nonlinear approaches apply.  The relaxation of these approximations along with the
identification of a deep connection between magnetodynamics and fluid
dynamics brings new perspectives on magnetism and reveals novel
physical regimes. Indeed,
nonlinear, dispersive physics are required to describe superfluids and
exotic structures such as solitons, quantized vortices, and dispersive
shock waves
(DSWs)~\cite{Hoefer2006,carusotto_bogoliubov-cerenkov_2006,fetter_rotating_2009,El2016},
as exemplified by Bose-Einstein condensates
(BECs)~\cite{Pethick2002,donley_dynamics_2001,Hoefer2006,el_oblique_2006,carusotto_bogoliubov-cerenkov_2006,cornish_formation_2006,gladush_radiation_2007,kevrekidis_emergent_2008,frantzeskakis_dark_2010,fetter_rotating_2009,kevrekidis_defocusing_2015,Demokritov2006,Qu2016,Congy2016}. To
obtain an analytical description of large-amplitude magnetic
textures, we introduce dispersive hydrodynamic
(DH) equations for a thin-film
ferromagnet.

This letter shows that the LL equation exactly maps into a DH system
of equations, without long-wavelength and low-frequency
restrictions. The conservative equations map to the Euler equations of
a compressible, isentropic fluid. The DH equations for a planar
ferromagnet admit spin-current-carrying, spatially periodic
magnetization textures termed ``uniform hydrodynamic states'' (UHSs),
providing a continuous and generalized description of spin
superflows~\cite{Konig2001,Sonin2010,Takei2014} up to small-amplitude
spin waves. Within the DH formulation, we prove that planar ferromagnets break Galilean invariance and elucidate their reference-frame-dependent dynamics by identifying the linear dispersion relation for spin waves propagating on top of a UHS background.  Such symmetry breaking at the level of linear spin waves is striking and fundamentally different from the non-trivial reference-frame-dependent dynamics of topological textures due to their inherent nonlinearity,  e.g., Walker breakdown
  for domain wall propagation~\cite{Beach2008} and core reversal in
  magnetic vortices~\cite{Yamada2007}. In this letter,
  we also show that static textures can break Galilean invariance for
  infinitesimal spin wave excitations that ride on a textured
  background. To emphasize this novel result, we suggest a Brillouin
  light scattering experimental test where broken Galilean invariance
  manifests itself as a spin-wave dispersion shift in the presence of
  a UHS.

We consider the nondimensionalized LL equation (see Supplementary
Material \cite{SuppMat})
\begin{equation}
  \label{eq:llnd}
  \frac{\partial \mathbf{m}}{\partial t} =
  -\mathbf{m}\times\mathbf{h}_\mathrm{eff}-\alpha\mathbf{m}\times\mathbf{m} 
  \times\mathbf{h}_\mathrm{eff}. 
\end{equation}
Damping is parametrized by the Gilbert constant $\alpha$,
$\mathbf{m}=\mathbf{M}/M_s = (m_x,m_y,m_z)$ is the magnetization
vector normalized to the saturation magnetization, and
$\mathbf{h}_\mathrm{eff}=\Delta\mathbf{m}-\sigma
m_z\hat{\mathbf{z}}+h_0 \hat{\mathbf{z}}$ is the normalized effective
field including: ferromagnetic exchange, $\Delta\mathbf{m}$; total
anisotropy determined by $\sigma=\mathrm{sgn}(M_s-H_k)$,
where $H_k$ is the perpendicular magnetic anisotropy field, such that
$\sigma=+1$ ($\sigma=-1$) represents a material with easy-plane
(perpendicular magnetic) anisotropy; and a perpendicular applied
field, $h_0$. This nondimensionalization of a two-dimensional (2D)
thin film provides a parameter-free description of materials with
planar or uniaxial anisotropy. We consider an idealized case where
in-plane magnetic anisotropy is negligible, i.e., its symmetry-breaking contribution only perturbs the leading order solution, similar to domain wall Brownian motion~\cite{Kim2015}.

Fluid-like variables are introduced using the canonical Hamiltonian
cylindrical transformation~\cite{papanicolaou_dynamics_1991}
\begin{equation}
  \label{eq:coord}
  n = m_z,\quad
  \mathbf{u} = - \nabla \Phi = -\nabla \left [
    \arctan{\left(m_y/m_x\right)} \right ],
\end{equation}
where $\Phi$ is the azimuthal phase angle. We identify $n$ ($|n| \leq
1$) as the \textit{longitudinal spin density} and $\mathbf{u}$ as the
\textit{fluid velocity}.  There are two important features of
Eq.~\eqref{eq:coord}. First, the flow is irrotational because the
velocity originates from a phase gradient, i.e., only quantized
circulation, such as a magnetic vortex~\cite{Sonin2010}, is
possible. Second, $\Phi$ is undefined when $n = \pm1$, corresponding to fluid \textit{vacuum}.

Utilizing the transformation~\eqref{eq:coord} and standard vector calculus identities, the LL
equation~\eqref{eq:llnd} can be exactly recast as two
DH equations~\cite{SuppMat}
\begin{subequations}
\label{eq:nudot}
\begin{eqnarray}
  \label{eq:ndot}
    \frac{\partial n}{\partial t} &=&
    \nabla\cdot\underbrace{\left[(1-n^2)\mathbf{u}\right]}_\text{spin
      current} + 
    \underbrace{\alpha(1-n^2)\frac{\partial \Phi}{\partial
        t}}_\text{spin relaxation} ,\\
  \label{eq:udot}
  \frac{\partial \mathbf{u}}{\partial t} &=&
  \nabla\underbrace{\left[(\sigma-|\mathbf{u}|^2)n
    \right]}_{\text{velocity flux}} 
  - \underbrace{\nabla\left[\frac{\Delta n}{1-n^2}+\frac{n |\nabla
        n|^2}{(1-n^2)^2}\right]}_\text{dispersion}\\  
  &&\underbrace{-\nabla
    h_0}_{\text{potential force}}
  +\underbrace{\alpha\nabla\left[\frac{1}{1-n^2}\nabla\cdot\left[(1-n^2) 
        \mathbf{u}\right]\right]}_\text{viscous loss}.\nonumber
\end{eqnarray}
\end{subequations}
Equation~\eqref{eq:ndot} is reminiscent of spin density continuity~\cite{Zhang2005} from which we identify the spin density
flux as the spin current
\begin{equation}
  \label{eq:2}
  \mathbf{J}_s = -(1-n^2)\mathbf{u} .
\end{equation}
Vacuum carries zero spin current. However,
maximal spin current is reached when $n=0$, identified as the
\textit{saturation density}. This implies that ferromagnetic textures ($\mathbf{u}\neq0)$ are better spin current conductors than small-amplitude spin waves~\cite{Hoffmann2013}. The hydrodynamic equivalents for the fluid velocity
Eq.~\eqref{eq:udot} are displayed.  When $n = |\nabla h_0| = 0$,
Eq.~\eqref{eq:udot} becomes $\partial \mathbf{u}/\partial t = \alpha
\nabla (\nabla\! \cdot\! \mathbf{u})$, a diffusion equation for the
velocity, hence $\alpha > 0$ acts similar to a viscosity. Previous
works~\cite{Halperin1969,Konig2001,Sonin2010,Takei2014} have neglected
exchange dispersion and nonlinearity in Eqs.~\eqref{eq:nudot} by
assuming the long-wavelength, near saturation density,
low-velocity limit, i.e., $|\nabla n| \ll 1$, $|n| \ll 1$, and
$|\mathbf{u}|^2 \ll 1$. As we show below, the full nonlinearity and exchange dispersion included in Eqs.~\eqref{eq:ndot} and
\eqref{eq:udot} are required to describe the existence and
stability regions of magnetic hydrodynamic states and broken Galilean invariance.

Insight can be gained from the homogeneous field $\nabla h_0 \to 0$,
conservative $\alpha \to 0$ limit, where Eqs.~\eqref{eq:nudot} become conservation laws for $n$ and $\mathbf{u}$. Notably, the
non-negative deviation from vacuum $(1-n^2)$, or \textit{fluid
  density}, is not conserved. A conservation
law for the momentum $\mathbf{p} = n \mathbf{u}$ can also be obtained
\begin{align}
  \label{eq:1}
  \frac{\partial \mathbf{p}}{\partial t} &= \nabla \cdot [(1 - n^2)
  \mathbf{u} \mathbf{u}^T ] + \nabla P(n,|\mathbf{u}|) + \nabla \cdot \left [ \frac{\nabla n \nabla n^T}{1-n^2}
  \right ]\nonumber \\
  &\quad - \nabla \left [ \frac{n \Delta n + \frac{1}{2}|\nabla
      n|^2}{1 - 
      n^2} + \frac{ n^2 |\nabla n |^2}{(1-n^2)^2} \right ], 
\end{align}
where the magnetic pressure is defined as
\begin{equation}
  \label{eq:5}
  P(n,|\mathbf{u}|) = \frac{1}{2}(1 + n^2)(\sigma - |\mathbf{u}|^2)
  -\sigma. 
\end{equation}
Equations~\eqref{eq:ndot} with $\alpha = 0$, and \eqref{eq:1} are
analogous to the time-reversed Euler equations expressing conservation of mass and
momentum for a 2D, compressible, isentropic fluid with a density- and velocity-dependent pressure $P$.

Additionally, the one-dimensional conservative limit of Eqs.~\eqref{eq:ndot} and \eqref{eq:udot} are exactly the equations describing polarization waves in two-component spinor Bose gases~\cite{Qu2016,Congy2016} and, in the near vacuum ($|n|\sim1$), long-wavelength,
and low-frequency regime, approximate
the mean field dynamics of a BEC~\cite{Pethick2002,satija_other_2011}. This
suggests that thin film ferromagnets are ripe for the exploration of
nonlinear structures observed in BECs, e.g.,
``Bosenovas''~\cite{donley_dynamics_2001,cornish_formation_2006} in
attractive ($\sigma=-1$); and dark
solitons~\cite{frantzeskakis_dark_2010},
vortices~\cite{fetter_rotating_2009}, and DSWs~\cite{Hoefer2006} in
repulsive ($\sigma=+1$) BECs.  Some of these structures have been
observed in uniaxial (dissipative
droplets~\cite{Mohseni2013,Chung2016,Macia2014}) and planar
(vortices~\cite{Mistral2008}) thin film ferromagnets. As we demonstrate,
hydrodynamic states are also supported.

Consider an ideal planar thin film ferromagnet ($\sigma=+1$)
and a homogeneous field.  Equations~\eqref{eq:nudot} admit a static
($\partial \Phi/\partial t=0$) solution \textit{with nonzero
    fluid velocity}, $\mathbf{u} = u\hat{\mathbf{x}}$, $|u|<1$,
$n=0$, and $h_0=0$. These are ground states known as spin-density
waves (SDWs)~\cite{Grunner1994} or soliton
lattices~\cite{Sonin2010} that minimize both exchange and anisotropy energies, i.e., any configuration with
$|u|<1$ is stable when $\mathbf{m}$ lies purely in-plane.  SDWs
exhibit a periodic structure that affords them topological stability
whereby the phase rotation can be unwound only by crossing a magnetic
pole ($|n|=1$)~\cite{Sonin2010,SuppMat}. For a non-zero field, $|h_0|
< 1-u^2$, SDWs are also supported for any ${|u| < 1}$ when
$n=h_0/(1-u^2)$ due to the longitudinal spin density relaxation effected by Eq.~\eqref{eq:ndot}. Such a relaxation maintains
$u$ and thus \textit{the topology of and finite spin current carried by} a
SDW. This property is identical to that of equilibrium transverse spin
currents in other magnetic textures including domain walls and
vortices~[Ref.~\onlinecite{Sonin2010}, Eq.~(4) in
Ref.~\onlinecite{Tserkovnyak2009}].

\begin{figure}
  \centering \includegraphics[width=5.5in]{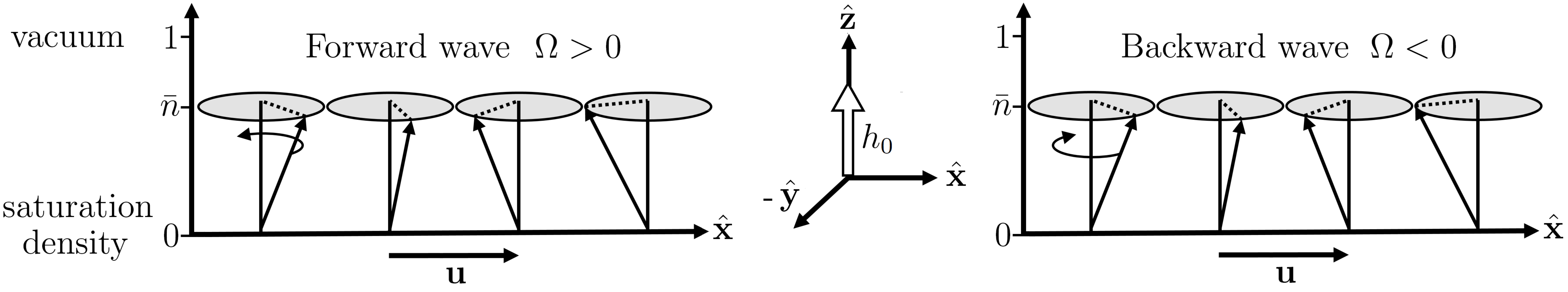}
  \caption{ \label{fig1} Schematic of a UHS. The longitudinal spin
    density is the vertical axis limited by vacuum ($|n|=1$) and
    saturation density ($n = 0$). Forward and backward wave
    conditions are determined by the sign of the frequency
    $\Omega$. }
\end{figure}

For no damping, Eqs.~\eqref{eq:nudot} admit dynamic solutions
parametrized by the constants $(\bar{n},\bar{u})$, where $|\bar{n}|\le
1$, $\mathbf{u}=\bar{u}\hat{\mathbf{x}}$, called \textit{uniform
  hydrodynamic state} (UHS). The fluid velocity $\bar{u}$ is the
wavenumber of the UHS whose frequency $\Omega=d\Phi/d t$ is
\begin{equation}
  \label{eq:8}
  \Omega(\bar{n},\bar{u})=-(1-\bar{u}^2)\bar{n}+h_0,
\end{equation}
obtained from the magnetic equivalent of Bernoulli's equation
${2P(\bar{n},|\bar{u}|) + \bar{u}^2 + \bar{n}(\Omega - h_0) =-\sigma}$
\cite{SuppMat}. Here, positive $\bar{u}$ implies clockwise
\textit{spatial} increase of the
  azimuthal phase $\Phi$ whereas negative $\Omega$ implies clockwise
\textit{temporal} precession about the $+\hat{\mathbf{z}}$ axis defining forward and backward wave conditions, schematically shown in
Fig.~\ref{fig1}. This is in contrast to magnetostatic forward and backward volume waves established by the relative direction between their wave vector and the external applied field.

The magnetization in a UHS can exhibit
large angle deviations from the $+\hat{\mathbf{z}}$ axis, making it a
strongly nonlinear texture. Near saturation density,
$|\bar{n}|\ll1$, a UHS limits to a spin
superflow~\cite{Konig2001,Sonin2010,Takei2014} whereas near vacuum,
$\bar{n} \sim \pm 1$, the UHS frequency Eq.~\eqref{eq:8} becomes the exchange spin-wave dispersion $\Omega \sim \pm \bar{u}^2 +
h_0 \mp 1$. Thus, a UHS is the generalization of periodic magnetic
textures from large (spin superflow) to small (spin-wave)
amplitudes. It is important to recognize that the ground state for the
UHS is a SDW, i.e., the ground state of planar ferromagnets is not
defined by a single orientation except for the vacuum state. In this
sense, the UHS is significantly different from the conventional theory
of spin waves based on the Holstein-Primakoff
transformation~\cite{Mattis2006} and their strongly nonlinear dynamics.~\cite{Lvov1994,Bertotti2008}.

Small-amplitude perturbations of a UHS can
  be regarded as spin waves with a generalized dispersion relation obtained from the linearization of Eqs.~\eqref{eq:ndot} and \eqref{eq:udot}
\begin{equation}
  \label{eq:dispersion}
  \omega_\pm(\mathbf{k},\mathbf{V}) = (2 \bar{n} \mathbf{u} -
  \mathbf{V}) \cdot \mathbf{k} \pm |\mathbf{k}|\sqrt{(1-\bar{n}^2)
    (1-\bar{u}^2)+|\mathbf{k}|^2}, 
\end{equation}
where $\mathbf{k}$ is the wave vector, and the velocity $\mathbf{V}$
reflects a Doppler shift, i.e., the velocity of an external observer
with respect to the UHS. The dispersion relation shows that
magnetic systems \textit{lack Galilean invariance}. In other words, an
observer velocity $\mathbf{V}\propto\mathbf{u}$ does not
generally result in a reference frame where the relative fluid
velocity is zero. Galilean invariance is recovered near vacuum with
dispersion
$\omega_\pm(\mathbf{k},\mathbf{V})=(2\mathbf{u}-\mathbf{V})\cdot\mathbf{k}\pm
|\mathbf{k}|^2$, i.e. exchange-mediated spin waves and the BEC
limit~\cite{Pethick2002,satija_other_2011}; and for spin superflow,
${\omega_\pm(\mathbf{k},\mathbf{V})=-\mathbf{V}\cdot\mathbf{k}\pm
  |\mathbf{k}|\sqrt{1+|\mathbf{k}|^2}}$. More importantly, the fluid
velocity $\bar{u}$ confers a spectral \emph{shift} in
  Eq.~\eqref{eq:dispersion} due to the UHS's intrinsically
chiral topology, similar to interfacial
Dzyaloshinskii-Moriya interaction~\cite{Nembach2015}.

\begin{figure}[t] 
  \centering \includegraphics[width=3.2in]{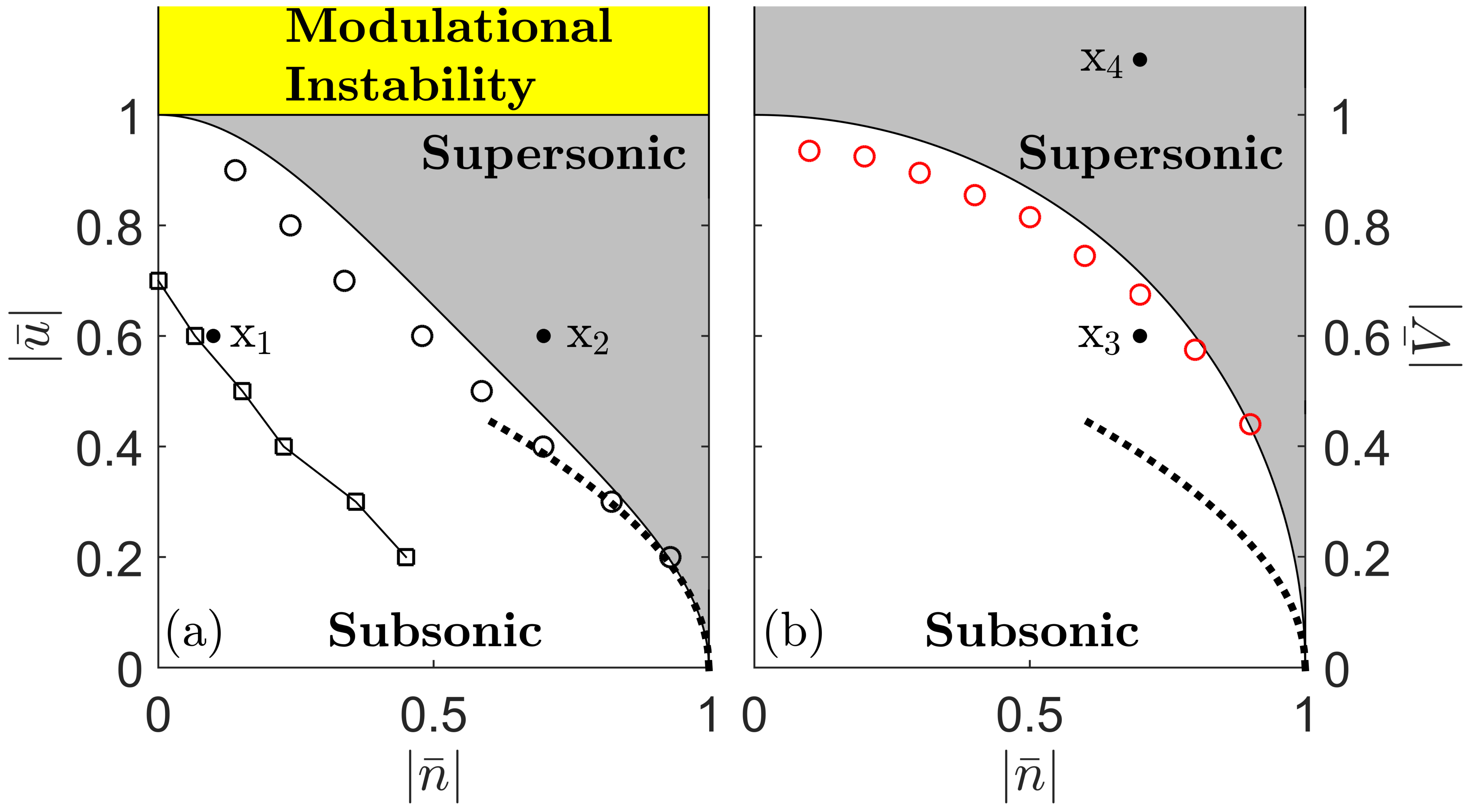}
  \caption{ \label{fig2} UHS phase diagram for (a) $\bar{V}=0$ and (b)
    $\bar{u}=0$ with subsonic (white), supersonic (gray), and
    modulationally unstable (yellow) regimes. Circles are numerical
    estimates of the sonic curves $\mathrm{M}_u=1$ and
    $\mathrm{M}_V=1$. The BEC regime sonic curve is dashed.
    Open squares represent the sonic curve of a 
    width $w=20$, thickness $\delta=1$ nanowire including non-local
    dipolar fields and $T=300$~K thermal field. Selected simulation
    conditions are denoted by x$_1$ to x$_4$.}
\end{figure}

The long wavelength limit of Eq.~\eqref{eq:dispersion} leads to
coincident spin-wave phase and group velocities, i.e., magnetic sound velocities
\begin{equation}
  \label{eq:7}
  s_\pm = 2\bar{n} \bar{u} + \bar{V} \pm \sqrt{(1 - \bar{n}^2)(1
    - \bar{u}^2)} .
\end{equation}
Here, we assume $\mathbf{V}$ collinear and opposite to $\mathbf{u}$
($\mathbf{V} = -\bar{V}\hat{\mathbf{x}}$). Subsonic flow occurs when
spin waves can propagate both forward and backward: $s_- < 0 < s_+$.
However, when $0 < s_- < s_+$, the flow is supersonic and some spin
waves are convected away. These conditions can be quantified in terms of
the Mach numbers $\mathrm{M}_{u}$, $\mathrm{M}_V$ when $\bar{V} = 0$,
$\bar{u} = 0$, respectively
\begin{equation}
  \label{eq:Mach}
  \mathrm{M}_{u} =
  |\bar{u}|\sqrt{\frac{1+3\bar{n}^2}{1-\bar{n}^2}}, \quad 
  \mathrm{M}_{V} = \frac{|\bar{V}|}{\sqrt{1-\bar{n}^2}} .
\end{equation}
For both, the flow is subsonic or laminar when $\mathrm{M} < 1$.  In
the supersonic regime, $\mathrm{M} > 1$, it is energetically favorable
to generate spin waves, thus leading to the Landau breakdown of
superfluid-like flow~\cite{landau_theory_1941}. The resulting phase
diagrams are shown in Fig.~\ref{fig2}. Interestingly,
Eq.~(\ref{eq:Mach}) predicts that $\mathrm{M}$ is independent of
$h_0$, implying that only the UHS longitudinal spin density and its
non-trivial topology, $\bar{u}$, set the supersonic transition, not the frequency $\Omega$. It must be noted
that broken Galilean invariance causes the Landau criterion concept
$\bar{u}<\mathrm{min}\left[s_\pm\right]$~\cite{Pethick2002} to give an
incorrect sonic curve.

A qualitatively distinct flow regime occurs when ${|\bar{u}| > 1}$ and
the sound velocities Eq.~\eqref{eq:7} are complex. This corresponds to
a change in the mathematical type of the long wavelength
Eqs.~\eqref{eq:nudot} from hyperbolic (wave-like) to elliptic
(potential-like).  Consequently, the UHS is unstable in the sense that
small fluctuations lead to drastic changes in its temporal evolution
or modulational instability (MI)~\cite{Whitham1974,zakharov_modulation_2009}.  Note that $|\bar{u}|<1$, $|\bar{V}| > 1$ does
not result in MI.

The aforementioned regimes were validated by performing
micromagnetic simulations with damping~\cite{Vansteenkiste2014}. We
simulate dynamics for an ideal Permalloy nanowire ($\mu_0M_s=1$~T) of
nondimensional width $w=20$ with transverse free spin boundary
  conditions and horizontal periodic boundary conditions (PBCs). We initialize with a SDW, include only local dipolar fields
(zero thickness limit), and set $\alpha = 0.01$. A homogeneous field
$h_0$ stabilizes the SDW at a specific $\bar{n}$ and a quantized
$\bar{u}$ that satisfies the PBC. This enables us to numerically probe
along a horizontal line in the phase diagram of Fig.~\ref{fig2}(a) by
implementing a slowly increasing $h_0$. By inserting a point defect (a
magnetic void), the SDW spontaneously generates spin waves when
$\bar{n}$ is large enough to cross the supersonic transition, leading
to a breakdown in the spatial homogeneity of the
SDW~\cite{SuppMat}. Due to the SDW's topology and the PBC, the change
in symmetry is accommodated by annihilating a single $2\pi$ phase rotation and reducing
$\bar{u}$ in a quantized fashion. Topologically, this is possible in
planar ferromagnets by crossing a magnetic pole, e.g., nucleating a
vortex, as shown in the Supplementary Video 1. This was also observed in wires of width $w=40$. The sonic curve
estimated this way is shown in Fig.~\ref{fig2}(a) by black circles,
demonstrating good agreement with $\mathrm{M}_u=1$. We attribute any
discrepancy to boundary and finite size
effects~\cite{frisch_transition_1992}, as further explored below.
\begin{figure}
  \centering \includegraphics[width=5.5in]{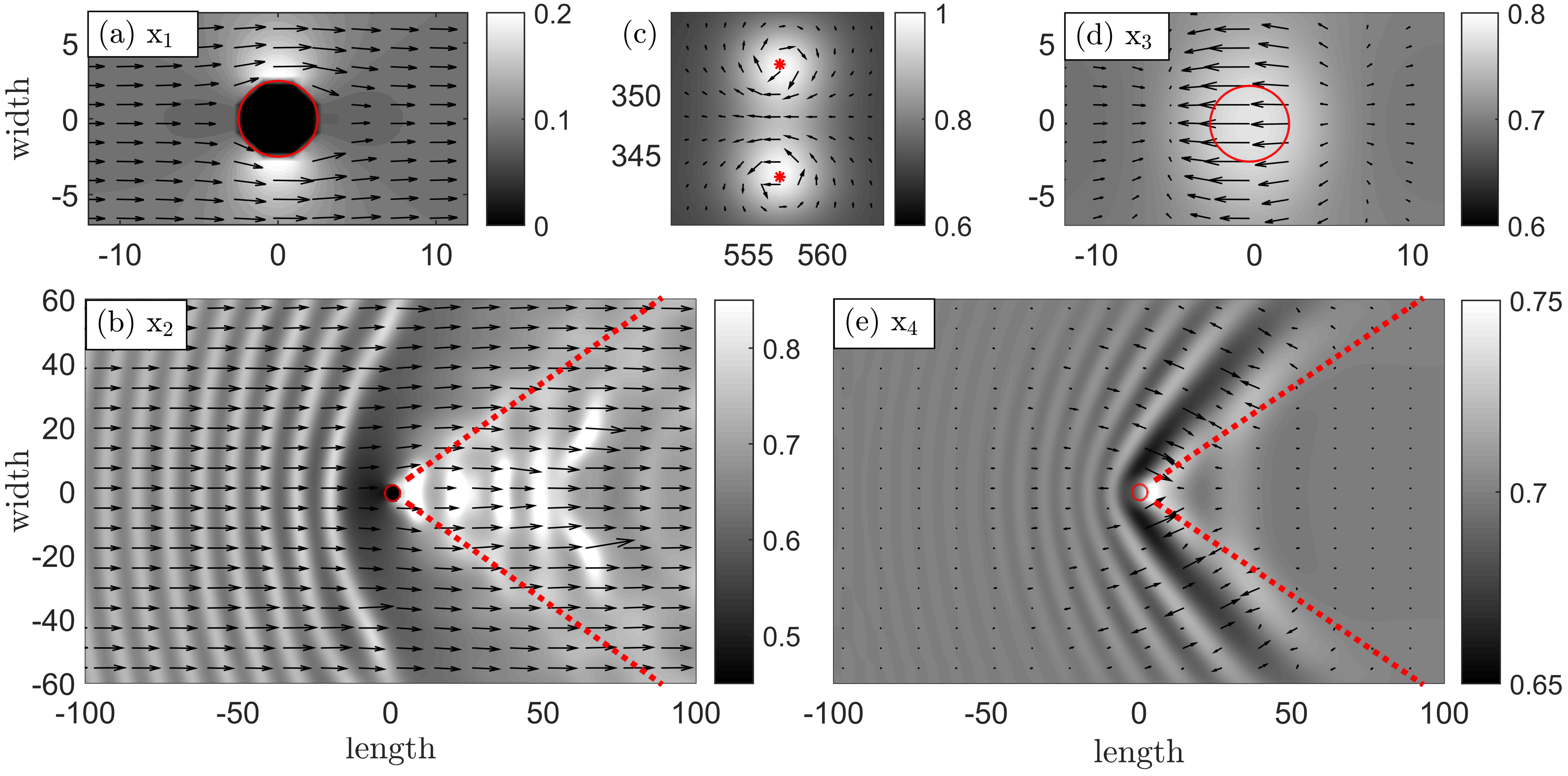}
  \caption{ \label{fig3} Snapshots of a (a), (b) SDW flowing past a stationary magnetic defect
    ($\bar{V}=0$) and; (d), (e) a homogeneous state subject to a moving, localized magnetic
    field ($\bar{V}\neq0$) with longitudinal spin density $n$
    (grayscale map) and velocity field $\mathbf{u}$ (arrows). The
    simulation region is much larger than what is visible. The defect
    or localized magnetic field position is shown by a red circle. For
    subsonic conditions, (a) and (d), the flow is static and
    laminar. In supersonic flow, (b) and (e), a Mach cone (dashed) and
    static wavefronts are observed. Propagating
    vortex-antivortex pairs with cores (asterisks) generated in (b)
    are shown in (c) far from the defect as opposite circulations of
    the velocity field (background $\bar{u}=0.6$ subtracted). }
\end{figure}

We use the same numerical method described above with the
  addition of thermal fluctuations and the symmetry-breaking non-local dipolar
fields to study the stability
of a SDW in a nanowire of nondimensional thickness $\delta=1$.  In this case, the SDW
topological structure completely collapses at the boundary shown in
Fig.~\ref{fig2}(a) by squares. In contrast to a recent
report where stable spin superflow was predicted only for nanowires shorter than the material exchange
length~\cite{Skarsvaag2015}, we observe stable SDWs over a wide range of parameters in phase space (Supplementary Video 2).

The supersonic transition in the moving frame is estimated by use of a
numerical method described elsewhere~\cite{Hoefer2012b}.  A moving,
perpendicular, localized, weak magnetic field spot with velocity
$\bar{V}$ is used to perturb a homogeneous state in the bias
field $h_0=\bar{n}$. The obtained sonic curve is in good agreement
with $\mathrm{M}_V = 1$, shown in Fig.~\ref{fig2}(b) by red
circles.

We now explore the effect of finite-sized obstacles on a UHS. As
observed in BECs, obstacles can generate vortices, wavefronts, and DSWs
in a fluid
flow~\cite{carusotto_bogoliubov-cerenkov_2006,fetter_rotating_2009,Hoefer2006}. Note that wavefronts, i.e., “spin-Cerenkov” radiation, were previously observed via micromagnetic simulations in homogeneous ($\bar{u}=0$),
  thick ferromagnets in the moving reference frame
  ($\bar{V}\neq0$)~\cite{Yan2013}. The wavefronts studied here are
  different, resulting from both moving ($\bar{u}=0$, $\bar{V}\neq0$) and static ($\bar{u}\neq0$, $\bar{V} = 0$) reference frames; yet another manifestation of broken Galilean invariance. We
illustrate these features with simulations where $\alpha = 0.01$ and local dipolar fields are included, shown
in Fig.~\ref{fig3} as a grayscale map and vector field
for $n$ and $\mathbf{u}$, respectively [see Supplement for the corresponding in-plane magnetization map]. First, we consider the
subsonic conditions x$_1$ of Fig.~\ref{fig2}(a) for a SDW in the static reference frame
($\bar{n},\bar{u})=(0.1,0.4$) with a magnetic defect within a circular
area of $\pi/\bar{u}$ in diameter. The static configuration in
Fig.~\ref{fig3}(a) is analogous to Bernoulli's principle for laminar
flow.

A different situation occurs at supersonic conditions x$_2$ ($\bar{n},\bar{u})=(0.7,0.6$), Fig.~\ref{fig3}(b). Here, the density develops a distinct Mach cone
(dashed), delimiting static wavefronts and the nucleation of propagating vortex-antivortex pairs, shown far from the
defect in Fig.~\ref{fig3}(c). In the moving reference frame, a homogeneous state is perturbed by a moving, weak,
localized field. Utilizing the subsonic conditions x$_3$
($\bar{n},\bar{u},\bar{V})=(0.7,0,0.6$), the flow is laminar, Fig.~\ref{fig3}(d)
[c.f.~supersonic x$_2$ in Fig.~\ref{fig2}(a)]. Wavefront radiation outside the Mach cone is observed for the supersonic
condition x$_4$ ($\bar{n},\bar{u},\bar{V})=(0.7,0,1.1$) in
Fig.~\ref{fig3}(e). However, the field spot amplitude is too weak to
excite vortex-antivortex pairs.  Animations are in Supplementary
Videos 3 to 6.

The MI regime for UHSs with $|\bar{u}|>1$ exhibits a violent
instability (see Supplementary Video 7). Notably, for
a uniaxial ferromagnet with $\sigma=-1$, MI is always operative. This
is consistent with the focusing of spin waves and the formation of
dissipative droplets in spin torque devices utilizing materials with
perpendicular magnetic
anisotropy~\cite{Hoefer2010,Mohseni2013,Macia2014,Chung2016}.

We now discuss an experimental test for the hydrodynamic predictions. As mentioned above, the dispersion relation Eq.~\eqref{eq:dispersion} features a spectral shift with non-zero fluid velocity. This shift can be observed e.g., by means of Brillouin light scattering (BLS), as already shown for Dzyaloshinskii-Moriya interactions~\cite{Nembach2015}. For a given fluid velocity, the magnitude of the shift, $2\bar{n}\bar{u}$, can be tuned by an externally applied field. Use of such tuning, in combination with BLS, will allow a direct test of the predicted breaking of Galilean invariance, insofar as the nonlinear properties of the dispersion relation Eq.~\eqref{eq:dispersion} can be quantitatively investigated. In particular, if one plots spin-wave frequency vs. wavenumber squared for both the Stokes and anti-Stokes BLS peaks in the short-wavelength limit, $|\mathbf{k}| \gg (1-\bar{n}^2)(1-\bar{u}^2)$, the modulus of the zero-wavenumber intercepts from linear regression \emph{will not be equal in the case of broken Galilean invariance}, in contrast to the case of Galilean invariance, where the intercepts should be equal.

In summary, the dispersive hydrodynamic (DH) formulation permits
  us to quantify the manner in which thin film ferromagnets lack
  Galilean invariance in the context of linear spin wave propagation
  on a dynamic UHS or static SDW background. The breaking of Galilean
invariance is often associated with relativistic phenomena wherein the
Lorentz transformation conjoins space-time into a single
  coordinate system, replacing the Galilean transformation. Instead,
the present case ultimately reflects the counterintuitive ability of
exchange-coupled, topological spin textures to support spin currents,
even in the static case. The predictions are robust to damping,
non-local dipolar fields, and finite temperatures for a large portion
of phase space. The exact representation of the LL equation in DH form
along with associated mathematical
tools~\cite{Whitham1974,kevrekidis_defocusing_2015,El2016} enables new magnetodynamic predictions and a frontier of
magnetic research, for example the observation of a Mach cone,
wavefronts, and vortex nucleation, suggesting the existence of
coherent structures such as oblique solitons and DSWs.

\begin{acknowledgments}
  The authors thank Leo Radzihovsky for beneficial
  discussions. E.I. acknowledges support from the Swedish Research
  Council, Reg. No. 637-2014-6863. M.A.H. partially supported by
  NSF CAREER DMS-1255422.
\end{acknowledgments}

\bibliographystyle{aipnum4-1}

\begin{thebibliography}{54}%
\makeatletter
\providecommand \@ifxundefined [1]{%
 \@ifx{#1\undefined}
}%
\providecommand \@ifnum [1]{%
 \ifnum #1\expandafter \@firstoftwo
 \else \expandafter \@secondoftwo
 \fi
}%
\providecommand \@ifx [1]{%
 \ifx #1\expandafter \@firstoftwo
 \else \expandafter \@secondoftwo
 \fi
}%
\providecommand \natexlab [1]{#1}%
\providecommand \enquote  [1]{``#1''}%
\providecommand \bibnamefont  [1]{#1}%
\providecommand \bibfnamefont [1]{#1}%
\providecommand \citenamefont [1]{#1}%
\providecommand \href@noop [0]{\@secondoftwo}%
\providecommand \href [0]{\begingroup \@sanitize@url \@href}%
\providecommand \@href[1]{\@@startlink{#1}\@@href}%
\providecommand \@@href[1]{\endgroup#1\@@endlink}%
\providecommand \@sanitize@url [0]{\catcode `\\12\catcode `\$12\catcode
  `\&12\catcode `\#12\catcode `\^12\catcode `\_12\catcode `\%12\relax}%
\providecommand \@@startlink[1]{}%
\providecommand \@@endlink[0]{}%
\providecommand \url  [0]{\begingroup\@sanitize@url \@url }%
\providecommand \@url [1]{\endgroup\@href {#1}{\urlprefix }}%
\providecommand \urlprefix  [0]{URL }%
\providecommand \Eprint [0]{\href }%
\providecommand \doibase [0]{http://dx.doi.org/}%
\providecommand \selectlanguage [0]{\@gobble}%
\providecommand \bibinfo  [0]{\@secondoftwo}%
\providecommand \bibfield  [0]{\@secondoftwo}%
\providecommand \translation [1]{[#1]}%
\providecommand \BibitemOpen [0]{}%
\providecommand \bibitemStop [0]{}%
\providecommand \bibitemNoStop [0]{.\EOS\space}%
\providecommand \EOS [0]{\spacefactor3000\relax}%
\providecommand \BibitemShut  [1]{\csname bibitem#1\endcsname}%
\let\auto@bib@innerbib\@empty
\bibitem [{\citenamefont {Berger}(1996)}]{Berger1996}%
  \BibitemOpen
  \bibfield  {author} {\bibinfo {author} {\bibfnamefont {L.}~\bibnamefont
  {Berger}},\ }\href {\doibase 10.1103/PhysRevB.54.9353} {\bibfield  {journal}
  {\bibinfo  {journal} {Phys. Rev. B}\ }\textbf {\bibinfo {volume} {54}},\
  \bibinfo {pages} {9353} (\bibinfo {year} {1996})}\BibitemShut {NoStop}%
\bibitem [{\citenamefont {Slonczewski}(1996)}]{Slonczewski1996}%
  \BibitemOpen
  \bibfield  {author} {\bibinfo {author} {\bibfnamefont {J.~C.}\ \bibnamefont
  {Slonczewski}},\ }\href {\doibase DOI: 10.1016/0304-8853(96)00062-5}
  {\bibfield  {journal} {\bibinfo  {journal} {J. Magn. Magn. Mater.}\ }\textbf
  {\bibinfo {volume} {159}},\ \bibinfo {pages} {L1 } (\bibinfo {year}
  {1996})}\BibitemShut {NoStop}%
\bibitem [{\citenamefont {Hals}\ and\ \citenamefont
  {Brataas}(2013)}]{Hals2013b}%
  \BibitemOpen
  \bibfield  {author} {\bibinfo {author} {\bibfnamefont {K.~M.~D.}\
  \bibnamefont {Hals}}\ and\ \bibinfo {author} {\bibfnamefont {A.}~\bibnamefont
  {Brataas}},\ }\href {\doibase 10.1103/PhysRevB.88.085423} {\bibfield
  {journal} {\bibinfo  {journal} {Phys. Rev. B}\ }\textbf {\bibinfo {volume}
  {88}},\ \bibinfo {pages} {085423} (\bibinfo {year} {2013})}\BibitemShut
  {NoStop}%
\bibitem [{\citenamefont {Hoefer}, \citenamefont {Silva},\ and\ \citenamefont
  {Keller}(2010)}]{Hoefer2010}%
  \BibitemOpen
  \bibfield  {author} {\bibinfo {author} {\bibfnamefont {M.~A.}\ \bibnamefont
  {Hoefer}}, \bibinfo {author} {\bibfnamefont {T.~J.}\ \bibnamefont {Silva}}, \
  and\ \bibinfo {author} {\bibfnamefont {M.~W.}\ \bibnamefont {Keller}},\
  }\href {\doibase 10.1103/PhysRevB.82.054432} {\bibfield  {journal} {\bibinfo
  {journal} {Phys. Rev. B}\ }\textbf {\bibinfo {volume} {82}},\ \bibinfo
  {pages} {054432} (\bibinfo {year} {2010})}\BibitemShut {NoStop}%
\bibitem [{\citenamefont {Mohseni}\ \emph {et~al.}(2013)\citenamefont
  {Mohseni}, \citenamefont {Sani}, \citenamefont {Persson}, \citenamefont
  {Nguyen}, \citenamefont {Chung}, \citenamefont {Pogoryelov}, \citenamefont
  {Muduli}, \citenamefont {Iacocca}, \citenamefont {Eklund}, \citenamefont
  {Dumas}, \citenamefont {Bonetti}, \citenamefont {Deac}, \citenamefont
  {Hoefer},\ and\ \citenamefont {\AA{}kerman}}]{Mohseni2013}%
  \BibitemOpen
  \bibfield  {author} {\bibinfo {author} {\bibfnamefont {S.~M.}\ \bibnamefont
  {Mohseni}}, \bibinfo {author} {\bibfnamefont {S.~R.}\ \bibnamefont {Sani}},
  \bibinfo {author} {\bibfnamefont {J.}~\bibnamefont {Persson}}, \bibinfo
  {author} {\bibfnamefont {T.~N.~A.}\ \bibnamefont {Nguyen}}, \bibinfo {author}
  {\bibfnamefont {S.}~\bibnamefont {Chung}}, \bibinfo {author} {\bibfnamefont
  {Y.}~\bibnamefont {Pogoryelov}}, \bibinfo {author} {\bibfnamefont {P.~K.}\
  \bibnamefont {Muduli}}, \bibinfo {author} {\bibfnamefont {E.}~\bibnamefont
  {Iacocca}}, \bibinfo {author} {\bibfnamefont {A.}~\bibnamefont {Eklund}},
  \bibinfo {author} {\bibfnamefont {R.~K.}\ \bibnamefont {Dumas}}, \bibinfo
  {author} {\bibfnamefont {S.}~\bibnamefont {Bonetti}}, \bibinfo {author}
  {\bibfnamefont {A.}~\bibnamefont {Deac}}, \bibinfo {author} {\bibfnamefont
  {M.~A.}\ \bibnamefont {Hoefer}}, \ and\ \bibinfo {author} {\bibfnamefont
  {J.}~\bibnamefont {\AA{}kerman}},\ }\href@noop {} {\bibfield  {journal}
  {\bibinfo  {journal} {Science}\ }\textbf {\bibinfo {volume} {339}},\ \bibinfo
  {pages} {1295} (\bibinfo {year} {2013})}\BibitemShut {NoStop}%
\bibitem [{\citenamefont {Maci\`{a}}, \citenamefont {Backes},\ and\
  \citenamefont {Kent}(2014)}]{Macia2014}%
  \BibitemOpen
  \bibfield  {author} {\bibinfo {author} {\bibfnamefont {F.}~\bibnamefont
  {Maci\`{a}}}, \bibinfo {author} {\bibfnamefont {D.}~\bibnamefont {Backes}}, \
  and\ \bibinfo {author} {\bibfnamefont {A.}~\bibnamefont {Kent}},\ }\href@noop
  {} {\bibfield  {journal} {\bibinfo  {journal} {Nature Nanotechnol}\ }\textbf
  {\bibinfo {volume} {10}},\ \bibinfo {pages} {1038} (\bibinfo {year}
  {2014})}\BibitemShut {NoStop}%
\bibitem [{\citenamefont {Chung}\ \emph {et~al.}(2016)\citenamefont {Chung},
  \citenamefont {Eklund}, \citenamefont {Iacocca}, \citenamefont {Mohseni},
  \citenamefont {Sani}, \citenamefont {Bookman}, \citenamefont {Hoefer},
  \citenamefont {Dumas},\ and\ \citenamefont {\AA{}kerman}}]{Chung2016}%
  \BibitemOpen
  \bibfield  {author} {\bibinfo {author} {\bibfnamefont {S.}~\bibnamefont
  {Chung}}, \bibinfo {author} {\bibfnamefont {A.}~\bibnamefont {Eklund}},
  \bibinfo {author} {\bibfnamefont {E.}~\bibnamefont {Iacocca}}, \bibinfo
  {author} {\bibfnamefont {S.}~\bibnamefont {Mohseni}}, \bibinfo {author}
  {\bibfnamefont {S.}~\bibnamefont {Sani}}, \bibinfo {author} {\bibfnamefont
  {L.}~\bibnamefont {Bookman}}, \bibinfo {author} {\bibfnamefont {M.~A.}\
  \bibnamefont {Hoefer}}, \bibinfo {author} {\bibfnamefont {R.}~\bibnamefont
  {Dumas}}, \ and\ \bibinfo {author} {\bibfnamefont {J.}~\bibnamefont
  {\AA{}kerman}},\ }\href@noop {} {\bibfield  {journal} {\bibinfo  {journal}
  {Nature Communications}\ }\textbf {\bibinfo {volume} {7}} (\bibinfo {year}
  {2016})}\BibitemShut {NoStop}%
\bibitem [{\citenamefont {Mistral}\ \emph {et~al.}(2008)\citenamefont
  {Mistral}, \citenamefont {van Kampen}, \citenamefont {Hrkac}, \citenamefont
  {Kim}, \citenamefont {Devolder}, \citenamefont {Crozat}, \citenamefont
  {Chappert}, \citenamefont {Lagae},\ and\ \citenamefont
  {Schrefl}}]{Mistral2008}%
  \BibitemOpen
  \bibfield  {author} {\bibinfo {author} {\bibfnamefont {Q.}~\bibnamefont
  {Mistral}}, \bibinfo {author} {\bibfnamefont {M.}~\bibnamefont {van Kampen}},
  \bibinfo {author} {\bibfnamefont {G.}~\bibnamefont {Hrkac}}, \bibinfo
  {author} {\bibfnamefont {J.-V.}\ \bibnamefont {Kim}}, \bibinfo {author}
  {\bibfnamefont {T.}~\bibnamefont {Devolder}}, \bibinfo {author}
  {\bibfnamefont {P.}~\bibnamefont {Crozat}}, \bibinfo {author} {\bibfnamefont
  {C.}~\bibnamefont {Chappert}}, \bibinfo {author} {\bibfnamefont
  {L.}~\bibnamefont {Lagae}}, \ and\ \bibinfo {author} {\bibfnamefont
  {T.}~\bibnamefont {Schrefl}},\ }\href {\doibase
  10.1103/PhysRevLett.100.257201} {\bibfield  {journal} {\bibinfo  {journal}
  {Phys. Rev. Lett.}\ }\textbf {\bibinfo {volume} {100}},\ \bibinfo {pages}
  {257201} (\bibinfo {year} {2008})}\BibitemShut {NoStop}%
\bibitem [{\citenamefont {Slavin}\ and\ \citenamefont
  {Tiberkevich}(2009)}]{Slavin2009}%
  \BibitemOpen
  \bibfield  {author} {\bibinfo {author} {\bibfnamefont {A.}~\bibnamefont
  {Slavin}}\ and\ \bibinfo {author} {\bibfnamefont {V.}~\bibnamefont
  {Tiberkevich}},\ }\href@noop {} {\bibfield  {journal} {\bibinfo  {journal}
  {Magnetics, IEEE Transactions on}\ }\textbf {\bibinfo {volume} {45}},\
  \bibinfo {pages} {1875 } (\bibinfo {year} {2009})}\BibitemShut {NoStop}%
\bibitem [{\citenamefont {L'vov}(1994)}]{Lvov1994}%
  \BibitemOpen
  \bibfield  {author} {\bibinfo {author} {\bibfnamefont {V.}~\bibnamefont
  {L'vov}},\ }\href@noop {} {\emph {\bibinfo {title} {Wave turbulence under
  parametric excitation}}}\ (\bibinfo  {publisher} {Springer},\ \bibinfo {year}
  {1994})\BibitemShut {NoStop}%
\bibitem [{\citenamefont {Bertotti}, \citenamefont {Mayergoyz},\ and\
  \citenamefont {Serpico}(2008)}]{Bertotti2008}%
  \BibitemOpen
  \bibfield  {author} {\bibinfo {author} {\bibfnamefont {G.}~\bibnamefont
  {Bertotti}}, \bibinfo {author} {\bibfnamefont {I.}~\bibnamefont {Mayergoyz}},
  \ and\ \bibinfo {author} {\bibfnamefont {C.}~\bibnamefont {Serpico}},\
  }\href@noop {} {\emph {\bibinfo {title} {Nonlinear magnetization dynamics in
  nanosystems}}},\ \bibinfo {edition} {first edition}\ ed.\ (\bibinfo
  {publisher} {Elsevier Science},\ \bibinfo {year} {2008})\BibitemShut
  {NoStop}%
\bibitem [{\citenamefont {Vansteenkiste}\ \emph {et~al.}(2014)\citenamefont
  {Vansteenkiste}, \citenamefont {Leliaert}, \citenamefont {Dvornik},
  \citenamefont {Helsen}, \citenamefont {Garcia-Sanchez},\ and\ \citenamefont
  {Van~Waeyenberge}}]{Vansteenkiste2014}%
  \BibitemOpen
  \bibfield  {author} {\bibinfo {author} {\bibfnamefont {A.}~\bibnamefont
  {Vansteenkiste}}, \bibinfo {author} {\bibfnamefont {J.}~\bibnamefont
  {Leliaert}}, \bibinfo {author} {\bibfnamefont {M.}~\bibnamefont {Dvornik}},
  \bibinfo {author} {\bibfnamefont {M.}~\bibnamefont {Helsen}}, \bibinfo
  {author} {\bibfnamefont {F.}~\bibnamefont {Garcia-Sanchez}}, \ and\ \bibinfo
  {author} {\bibfnamefont {B.}~\bibnamefont {Van~Waeyenberge}},\ }\href@noop {}
  {\bibfield  {journal} {\bibinfo  {journal} {AIP Advances}\ }\textbf {\bibinfo
  {volume} {4}},\ \bibinfo {eid} {107133} (\bibinfo {year} {2014})}\BibitemShut
  {NoStop}%
\bibitem [{\citenamefont {Halperin}\ and\ \citenamefont
  {Hohenberg}(1969)}]{Halperin1969}%
  \BibitemOpen
  \bibfield  {author} {\bibinfo {author} {\bibfnamefont {B.}~\bibnamefont
  {Halperin}}\ and\ \bibinfo {author} {\bibfnamefont {P.}~\bibnamefont
  {Hohenberg}},\ }\href@noop {} {\bibfield  {journal} {\bibinfo  {journal}
  {Physical Review}\ }\textbf {\bibinfo {volume} {188}},\ \bibinfo {pages}
  {898} (\bibinfo {year} {1969})}\BibitemShut {NoStop}%
\bibitem [{\citenamefont {K\"onig}, \citenamefont {B\o{}nsager},\ and\
  \citenamefont {MacDonald}(2001)}]{Konig2001}%
  \BibitemOpen
  \bibfield  {author} {\bibinfo {author} {\bibfnamefont {J.}~\bibnamefont
  {K\"onig}}, \bibinfo {author} {\bibfnamefont {M.~C.}\ \bibnamefont
  {B\o{}nsager}}, \ and\ \bibinfo {author} {\bibfnamefont {A.~H.}\ \bibnamefont
  {MacDonald}},\ }\href {\doibase 10.1103/PhysRevLett.87.187202} {\bibfield
  {journal} {\bibinfo  {journal} {Phys. Rev. Lett.}\ }\textbf {\bibinfo
  {volume} {87}},\ \bibinfo {pages} {187202} (\bibinfo {year}
  {2001})}\BibitemShut {NoStop}%
\bibitem [{\citenamefont {Sonin}(2010)}]{Sonin2010}%
  \BibitemOpen
  \bibfield  {author} {\bibinfo {author} {\bibfnamefont {E.~B.}\ \bibnamefont
  {Sonin}},\ }\href@noop {} {\bibfield  {journal} {\bibinfo  {journal}
  {Advances in Physics}\ }\textbf {\bibinfo {volume} {59}},\ \bibinfo {pages}
  {181 } (\bibinfo {year} {2010})}\BibitemShut {NoStop}%
\bibitem [{\citenamefont {Takei}\ and\ \citenamefont
  {Tserkovnyak}(2014)}]{Takei2014}%
  \BibitemOpen
  \bibfield  {author} {\bibinfo {author} {\bibfnamefont {S.}~\bibnamefont
  {Takei}}\ and\ \bibinfo {author} {\bibfnamefont {Y.}~\bibnamefont
  {Tserkovnyak}},\ }\href@noop {} {\bibfield  {journal} {\bibinfo  {journal}
  {Phys. Rev. Lett.}\ }\textbf {\bibinfo {volume} {112}},\ \bibinfo {pages}
  {227201} (\bibinfo {year} {2014})}\BibitemShut {NoStop}%
\bibitem [{\citenamefont {Chen}\ \emph {et~al.}(2014)\citenamefont {Chen},
  \citenamefont {Kent}, \citenamefont {MacDonald},\ and\ \citenamefont
  {Sodemann}}]{Chen2014}%
  \BibitemOpen
  \bibfield  {author} {\bibinfo {author} {\bibfnamefont {H.}~\bibnamefont
  {Chen}}, \bibinfo {author} {\bibfnamefont {A.~D.}\ \bibnamefont {Kent}},
  \bibinfo {author} {\bibfnamefont {A.~H.}\ \bibnamefont {MacDonald}}, \ and\
  \bibinfo {author} {\bibfnamefont {I.}~\bibnamefont {Sodemann}},\ }\href
  {\doibase 10.1103/PhysRevB.90.220401} {\bibfield  {journal} {\bibinfo
  {journal} {Phys. Rev. B}\ }\textbf {\bibinfo {volume} {90}},\ \bibinfo
  {pages} {220401} (\bibinfo {year} {2014})}\BibitemShut {NoStop}%
\bibitem [{\citenamefont {Skarsv\aa{}g}, \citenamefont {Holmqvist},\ and\
  \citenamefont {Brataas}(2015)}]{Skarsvaag2015}%
  \BibitemOpen
  \bibfield  {author} {\bibinfo {author} {\bibfnamefont {H.}~\bibnamefont
  {Skarsv\aa{}g}}, \bibinfo {author} {\bibfnamefont {C.}~\bibnamefont
  {Holmqvist}}, \ and\ \bibinfo {author} {\bibfnamefont {A.}~\bibnamefont
  {Brataas}},\ }\href {\doibase 10.1103/PhysRevLett.115.237201} {\bibfield
  {journal} {\bibinfo  {journal} {Phys. Rev. Lett.}\ }\textbf {\bibinfo
  {volume} {115}},\ \bibinfo {pages} {237201} (\bibinfo {year}
  {2015})}\BibitemShut {NoStop}%
\bibitem [{\citenamefont {Flebus}\ \emph {et~al.}(2016)\citenamefont {Flebus},
  \citenamefont {Bender}, \citenamefont {Tserkovnyak},\ and\ \citenamefont
  {Duine}}]{Flebus2016}%
  \BibitemOpen
  \bibfield  {author} {\bibinfo {author} {\bibfnamefont {B.}~\bibnamefont
  {Flebus}}, \bibinfo {author} {\bibfnamefont {S.~A.}\ \bibnamefont {Bender}},
  \bibinfo {author} {\bibfnamefont {Y.}~\bibnamefont {Tserkovnyak}}, \ and\
  \bibinfo {author} {\bibfnamefont {R.~A.}\ \bibnamefont {Duine}},\ }\href
  {\doibase 10.1103/PhysRevLett.116.117201} {\bibfield  {journal} {\bibinfo
  {journal} {Phys. Rev. Lett.}\ }\textbf {\bibinfo {volume} {116}},\ \bibinfo
  {pages} {117201} (\bibinfo {year} {2016})}\BibitemShut {NoStop}%
\bibitem [{\citenamefont {Hoefer}\ \emph {et~al.}(2006)\citenamefont {Hoefer},
  \citenamefont {Ablowitz}, \citenamefont {Coddington}, \citenamefont
  {Cornell}, \citenamefont {Engels},\ and\ \citenamefont
  {Schweikhard}}]{Hoefer2006}%
  \BibitemOpen
  \bibfield  {author} {\bibinfo {author} {\bibfnamefont {M.~A.}\ \bibnamefont
  {Hoefer}}, \bibinfo {author} {\bibfnamefont {M.~J.}\ \bibnamefont
  {Ablowitz}}, \bibinfo {author} {\bibfnamefont {I.}~\bibnamefont
  {Coddington}}, \bibinfo {author} {\bibfnamefont {E.~A.}\ \bibnamefont
  {Cornell}}, \bibinfo {author} {\bibfnamefont {P.}~\bibnamefont {Engels}}, \
  and\ \bibinfo {author} {\bibfnamefont {V.}~\bibnamefont {Schweikhard}},\
  }\href {\doibase 10.1103/PhysRevA.74.023623} {\bibfield  {journal} {\bibinfo
  {journal} {Phys. Rev. A}\ }\textbf {\bibinfo {volume} {74}},\ \bibinfo
  {pages} {023623} (\bibinfo {year} {2006})}\BibitemShut {NoStop}%
\bibitem [{\citenamefont {Carusotto}\ \emph {et~al.}(2006)\citenamefont
  {Carusotto}, \citenamefont {Hu}, \citenamefont {Collins},\ and\ \citenamefont
  {Smerzi}}]{carusotto_bogoliubov-cerenkov_2006}%
  \BibitemOpen
  \bibfield  {author} {\bibinfo {author} {\bibfnamefont {I.}~\bibnamefont
  {Carusotto}}, \bibinfo {author} {\bibfnamefont {S.~X.}\ \bibnamefont {Hu}},
  \bibinfo {author} {\bibfnamefont {L.~A.}\ \bibnamefont {Collins}}, \ and\
  \bibinfo {author} {\bibfnamefont {A.}~\bibnamefont {Smerzi}},\ }\href@noop {}
  {\bibfield  {journal} {\bibinfo  {journal} {Phys. Rev. Lett.}\ }\textbf
  {\bibinfo {volume} {97}},\ \bibinfo {pages} {260403} (\bibinfo {year}
  {2006})}\BibitemShut {NoStop}%
\bibitem [{\citenamefont {Fetter}(2009)}]{fetter_rotating_2009}%
  \BibitemOpen
  \bibfield  {author} {\bibinfo {author} {\bibfnamefont {A.~L.}\ \bibnamefont
  {Fetter}},\ }\href@noop {} {\bibfield  {journal} {\bibinfo  {journal}
  {Reviews of Modern Physics}\ }\textbf {\bibinfo {volume} {81}},\ \bibinfo
  {pages} {647} (\bibinfo {year} {2009})}\BibitemShut {NoStop}%
\bibitem [{\citenamefont {El}\ and\ \citenamefont {Hoefer}(2016)}]{El2016}%
  \BibitemOpen
  \bibfield  {author} {\bibinfo {author} {\bibfnamefont {G.}~\bibnamefont
  {El}}\ and\ \bibinfo {author} {\bibfnamefont {M.}~\bibnamefont {Hoefer}},\
  }\href@noop {} {\bibfield  {journal} {\bibinfo  {journal} {Physica D, to
  appear}\ } (\bibinfo {year} {2016})}\BibitemShut {NoStop}%
\bibitem [{\citenamefont {Pethick}\ and\ \citenamefont
  {Smith}(2002)}]{Pethick2002}%
  \BibitemOpen
  \bibfield  {author} {\bibinfo {author} {\bibfnamefont {C.}~\bibnamefont
  {Pethick}}\ and\ \bibinfo {author} {\bibfnamefont {H.}~\bibnamefont
  {Smith}},\ }\href@noop {} {\emph {\bibinfo {title} {Bose-Einstein
  condensation in dilute gasses}}}\ (\bibinfo  {publisher} {Cambridge
  University Press},\ \bibinfo {year} {2002})\BibitemShut {NoStop}%
\bibitem [{\citenamefont {Donley}\ \emph {et~al.}(2001)\citenamefont {Donley},
  \citenamefont {Claussen}, \citenamefont {Cornish}, \citenamefont {Roberts},
  \citenamefont {Cornell},\ and\ \citenamefont
  {Wieman}}]{donley_dynamics_2001}%
  \BibitemOpen
  \bibfield  {author} {\bibinfo {author} {\bibfnamefont {E.~A.}\ \bibnamefont
  {Donley}}, \bibinfo {author} {\bibfnamefont {N.~R.}\ \bibnamefont
  {Claussen}}, \bibinfo {author} {\bibfnamefont {S.~L.}\ \bibnamefont
  {Cornish}}, \bibinfo {author} {\bibfnamefont {J.~L.}\ \bibnamefont
  {Roberts}}, \bibinfo {author} {\bibfnamefont {E.~A.}\ \bibnamefont
  {Cornell}}, \ and\ \bibinfo {author} {\bibfnamefont {C.~E.}\ \bibnamefont
  {Wieman}},\ }\href@noop {} {\bibfield  {journal} {\bibinfo  {journal}
  {Nature}\ }\textbf {\bibinfo {volume} {412}},\ \bibinfo {pages} {295}
  (\bibinfo {year} {2001})}\BibitemShut {NoStop}%
\bibitem [{\citenamefont {El}, \citenamefont {Gammal},\ and\ \citenamefont
  {Kamchatnov}(2006)}]{el_oblique_2006}%
  \BibitemOpen
  \bibfield  {author} {\bibinfo {author} {\bibfnamefont {G.~A.}\ \bibnamefont
  {El}}, \bibinfo {author} {\bibfnamefont {A.}~\bibnamefont {Gammal}}, \ and\
  \bibinfo {author} {\bibfnamefont {A.~M.}\ \bibnamefont {Kamchatnov}},\
  }\href@noop {} {\bibfield  {journal} {\bibinfo  {journal} {Phys. Rev. Lett.}\
  }\textbf {\bibinfo {volume} {97}},\ \bibinfo {pages} {180405} (\bibinfo
  {year} {2006})}\BibitemShut {NoStop}%
\bibitem [{\citenamefont {Cornish}, \citenamefont {Thompson},\ and\
  \citenamefont {Wieman}(2006)}]{cornish_formation_2006}%
  \BibitemOpen
  \bibfield  {author} {\bibinfo {author} {\bibfnamefont {S.~L.}\ \bibnamefont
  {Cornish}}, \bibinfo {author} {\bibfnamefont {S.~T.}\ \bibnamefont
  {Thompson}}, \ and\ \bibinfo {author} {\bibfnamefont {C.~E.}\ \bibnamefont
  {Wieman}},\ }\href@noop {} {\bibfield  {journal} {\bibinfo  {journal}
  {Physical Review Letters}\ }\textbf {\bibinfo {volume} {96}},\ \bibinfo
  {pages} {170401} (\bibinfo {year} {2006})}\BibitemShut {NoStop}%
\bibitem [{\citenamefont {Gladush}\ \emph {et~al.}(2007)\citenamefont
  {Gladush}, \citenamefont {A.~El}, \citenamefont {Gammal},\ and\ \citenamefont
  {Kamchatnov}}]{gladush_radiation_2007}%
  \BibitemOpen
  \bibfield  {author} {\bibinfo {author} {\bibfnamefont {Y.~G.}\ \bibnamefont
  {Gladush}}, \bibinfo {author} {\bibfnamefont {G.}~\bibnamefont {A.~El}},
  \bibinfo {author} {\bibfnamefont {A.}~\bibnamefont {Gammal}}, \ and\ \bibinfo
  {author} {\bibfnamefont {A.~M.}\ \bibnamefont {Kamchatnov}},\ }\href@noop {}
  {\bibfield  {journal} {\bibinfo  {journal} {Phys. Rev. A}\ }\textbf {\bibinfo
  {volume} {75}},\ \bibinfo {pages} {033619} (\bibinfo {year}
  {2007})}\BibitemShut {NoStop}%
\bibitem [{\citenamefont {Kevrekidis}, \citenamefont {Frantzeskakis},\ and\
  \citenamefont {Carretero-Gonz\`{a}lez}(2008)}]{kevrekidis_emergent_2008}%
  \BibitemOpen
  \bibfield  {author} {\bibinfo {author} {\bibfnamefont {P.~G.}\ \bibnamefont
  {Kevrekidis}}, \bibinfo {author} {\bibfnamefont {D.~J.}\ \bibnamefont
  {Frantzeskakis}}, \ and\ \bibinfo {author} {\bibfnamefont {R.}~\bibnamefont
  {Carretero-Gonz\`{a}lez}},\ }\href@noop {} {\emph {\bibinfo {title} {Emergent
  nonlinear phenomena in {Bose}-{Einstein} condensates: theory and
  experiment}}}\ (\bibinfo  {publisher} {Springer},\ \bibinfo {address}
  {Berlin},\ \bibinfo {year} {2008})\BibitemShut {NoStop}%
\bibitem [{\citenamefont {Frantzeskakis}(2010)}]{frantzeskakis_dark_2010}%
  \BibitemOpen
  \bibfield  {author} {\bibinfo {author} {\bibfnamefont {D.~J.}\ \bibnamefont
  {Frantzeskakis}},\ }\href@noop {} {\bibfield  {journal} {\bibinfo  {journal}
  {Journal of Physics A: Mathematical and Theoretical}\ }\textbf {\bibinfo
  {volume} {43}},\ \bibinfo {pages} {213001} (\bibinfo {year}
  {2010})}\BibitemShut {NoStop}%
\bibitem [{\citenamefont {Kevrekidis}, \citenamefont {Frantzeskakis},\ and\
  \citenamefont {Carretero-Gonz\'{a}lez}(2015)}]{kevrekidis_defocusing_2015}%
  \BibitemOpen
  \bibfield  {author} {\bibinfo {author} {\bibfnamefont {P.~G.}\ \bibnamefont
  {Kevrekidis}}, \bibinfo {author} {\bibfnamefont {D.~J.}\ \bibnamefont
  {Frantzeskakis}}, \ and\ \bibinfo {author} {\bibfnamefont {R.}~\bibnamefont
  {Carretero-Gonz\'{a}lez}},\ }\href@noop {} {\emph {\bibinfo {title} {The
  {Defocusing} {Nonlinear} {Schr\"{o}dinger} {Equation}}}}\ (\bibinfo
  {publisher} {SIAM},\ \bibinfo {address} {Philadelphia},\ \bibinfo {year}
  {2015})\BibitemShut {NoStop}%
\bibitem [{\citenamefont {Demokritov}\ \emph {et~al.}(2006)\citenamefont
  {Demokritov}, \citenamefont {Demidov}, \citenamefont {Dzyapko}, \citenamefont
  {Melkov}, \citenamefont {Serga}, \citenamefont {Hillebrands},\ and\
  \citenamefont {Slavin}}]{Demokritov2006}%
  \BibitemOpen
  \bibfield  {author} {\bibinfo {author} {\bibfnamefont {S.}~\bibnamefont
  {Demokritov}}, \bibinfo {author} {\bibfnamefont {V.}~\bibnamefont {Demidov}},
  \bibinfo {author} {\bibfnamefont {O.}~\bibnamefont {Dzyapko}}, \bibinfo
  {author} {\bibfnamefont {G.~A.}\ \bibnamefont {Melkov}}, \bibinfo {author}
  {\bibfnamefont {A.~A.}\ \bibnamefont {Serga}}, \bibinfo {author}
  {\bibfnamefont {B.}~\bibnamefont {Hillebrands}}, \ and\ \bibinfo {author}
  {\bibfnamefont {A.}~\bibnamefont {Slavin}},\ }\href@noop {} {\bibfield
  {journal} {\bibinfo  {journal} {Nature}\ }\textbf {\bibinfo {volume} {443}},\
  \bibinfo {pages} {430} (\bibinfo {year} {2006})}\BibitemShut {NoStop}%
\bibitem [{\citenamefont {Qu}, \citenamefont {Pitaevskii},\ and\ \citenamefont
  {Stringari}(2016)}]{Qu2016}%
  \BibitemOpen
  \bibfield  {author} {\bibinfo {author} {\bibfnamefont {C.}~\bibnamefont
  {Qu}}, \bibinfo {author} {\bibfnamefont {L.~P.}\ \bibnamefont {Pitaevskii}},
  \ and\ \bibinfo {author} {\bibfnamefont {S.}~\bibnamefont {Stringari}},\
  }\href {\doibase 10.1103/PhysRevLett.116.160402} {\bibfield  {journal}
  {\bibinfo  {journal} {Phys. Rev. Lett.}\ }\textbf {\bibinfo {volume} {116}},\
  \bibinfo {pages} {160402} (\bibinfo {year} {2016})}\BibitemShut {NoStop}%
\bibitem [{\citenamefont {Congy}, \citenamefont {Kamchatnov},\ and\
  \citenamefont {Pavloff}(2016)}]{Congy2016}%
  \BibitemOpen
  \bibfield  {author} {\bibinfo {author} {\bibfnamefont {T.}~\bibnamefont
  {Congy}}, \bibinfo {author} {\bibfnamefont {A.~M.}\ \bibnamefont
  {Kamchatnov}}, \ and\ \bibinfo {author} {\bibfnamefont {N.}~\bibnamefont
  {Pavloff}},\ }\href@noop {} {\bibfield  {journal} {\bibinfo  {journal}
  {arXiv:1607.08760}\ } (\bibinfo {year} {2016})}\BibitemShut {NoStop}%
\bibitem [{\citenamefont {Beach}, \citenamefont {Tsoi},\ and\ \citenamefont
  {Erskine}(2008)}]{Beach2008}%
  \BibitemOpen
  \bibfield  {author} {\bibinfo {author} {\bibfnamefont {G.}~\bibnamefont
  {Beach}}, \bibinfo {author} {\bibfnamefont {M.}~\bibnamefont {Tsoi}}, \ and\
  \bibinfo {author} {\bibfnamefont {J.}~\bibnamefont {Erskine}},\ }\href@noop
  {} {\bibfield  {journal} {\bibinfo  {journal} {Journal of Magnetism and
  Magnetic Materials}\ }\textbf {\bibinfo {volume} {320}},\ \bibinfo {pages}
  {1272 } (\bibinfo {year} {2008})}\BibitemShut {NoStop}%
\bibitem [{\citenamefont {Yamada}\ \emph {et~al.}(2007)\citenamefont {Yamada},
  \citenamefont {Kasai}, \citenamefont {Nakatani}, \citenamefont {Kobayashi},
  \citenamefont {Kohno}, \citenamefont {Thiaville},\ and\ \citenamefont
  {Ono}}]{Yamada2007}%
  \BibitemOpen
  \bibfield  {author} {\bibinfo {author} {\bibfnamefont {K.}~\bibnamefont
  {Yamada}}, \bibinfo {author} {\bibfnamefont {S.}~\bibnamefont {Kasai}},
  \bibinfo {author} {\bibfnamefont {Y.}~\bibnamefont {Nakatani}}, \bibinfo
  {author} {\bibfnamefont {K.}~\bibnamefont {Kobayashi}}, \bibinfo {author}
  {\bibfnamefont {H.}~\bibnamefont {Kohno}}, \bibinfo {author} {\bibfnamefont
  {A.}~\bibnamefont {Thiaville}}, \ and\ \bibinfo {author} {\bibfnamefont
  {T.}~\bibnamefont {Ono}},\ }\href@noop {} {\bibfield  {journal} {\bibinfo
  {journal} {Nature}\ }\textbf {\bibinfo {volume} {6}},\ \bibinfo {pages} {269}
  (\bibinfo {year} {2007})}\BibitemShut {NoStop}%
\bibitem [{Sup()}]{SuppMat}%
  \BibitemOpen
  \href@noop {} {\bibinfo  {journal} {See Supplementary Material [url], which
  includes Refs. [53-54]}\ }\BibitemShut {NoStop}%
\bibitem [{\citenamefont {Kim}, \citenamefont {Takei},\ and\ \citenamefont
  {Tserkovnyak}(2015)}]{Kim2015}%
  \BibitemOpen
\bibfield  {journal} {  }\bibfield  {author} {\bibinfo {author} {\bibfnamefont
  {S.~K.}\ \bibnamefont {Kim}}, \bibinfo {author} {\bibfnamefont
  {S.}~\bibnamefont {Takei}}, \ and\ \bibinfo {author} {\bibfnamefont
  {Y.}~\bibnamefont {Tserkovnyak}},\ }\href {\doibase
  10.1103/PhysRevB.92.220409} {\bibfield  {journal} {\bibinfo  {journal} {Phys.
  Rev. B}\ }\textbf {\bibinfo {volume} {92}},\ \bibinfo {pages} {220409}
  (\bibinfo {year} {2015})}\BibitemShut {NoStop}%
\bibitem [{\citenamefont {Papanicolaou}\ and\ \citenamefont
  {Tomaras}(1991)}]{papanicolaou_dynamics_1991}%
  \BibitemOpen
  \bibfield  {author} {\bibinfo {author} {\bibfnamefont {N.}~\bibnamefont
  {Papanicolaou}}\ and\ \bibinfo {author} {\bibfnamefont {T.}~\bibnamefont
  {Tomaras}},\ }\href {\doibase 10.1016/0550-3213(91)90410-Y} {\bibfield
  {journal} {\bibinfo  {journal} {Nuclear Physics B}\ }\textbf {\bibinfo
  {volume} {360}},\ \bibinfo {pages} {425} (\bibinfo {year}
  {1991})}\BibitemShut {NoStop}%
\bibitem [{\citenamefont {Zhang}\ and\ \citenamefont {Yang}(2005)}]{Zhang2005}%
  \BibitemOpen
  \bibfield  {author} {\bibinfo {author} {\bibfnamefont {S.}~\bibnamefont
  {Zhang}}\ and\ \bibinfo {author} {\bibfnamefont {Z.}~\bibnamefont {Yang}},\
  }\href@noop {} {\bibfield  {journal} {\bibinfo  {journal} {Phys. Rev. Lett.}\
  }\textbf {\bibinfo {volume} {94}},\ \bibinfo {pages} {066602} (\bibinfo
  {year} {2005})}\BibitemShut {NoStop}%
\bibitem [{\citenamefont {Hoffmann}(2013)}]{Hoffmann2013}%
  \BibitemOpen
  \bibfield  {author} {\bibinfo {author} {\bibfnamefont {A.}~\bibnamefont
  {Hoffmann}},\ }\href@noop {} {\bibfield  {journal} {\bibinfo  {journal} {IEEE
  Advances on magnetics}\ }\textbf {\bibinfo {volume} {49}},\ \bibinfo {pages}
  {5172} (\bibinfo {year} {2013})}\BibitemShut {NoStop}%
\bibitem [{\citenamefont {Satija}\ and\ \citenamefont
  {Balakrishnan}(2011)}]{satija_other_2011}%
  \BibitemOpen
  \bibfield  {author} {\bibinfo {author} {\bibfnamefont {I.~I.}\ \bibnamefont
  {Satija}}\ and\ \bibinfo {author} {\bibfnamefont {R.}~\bibnamefont
  {Balakrishnan}},\ }\href@noop {} {\bibfield  {journal} {\bibinfo  {journal}
  {Phys. Lett. A}\ }\textbf {\bibinfo {volume} {375}},\ \bibinfo {pages} {517}
  (\bibinfo {year} {2011})}\BibitemShut {NoStop}%
\bibitem [{\citenamefont {Gr\"uner}(1994)}]{Grunner1994}%
  \BibitemOpen
  \bibfield  {author} {\bibinfo {author} {\bibfnamefont {G.}~\bibnamefont
  {Gr\"uner}},\ }\href {\doibase 10.1103/RevModPhys.66.1} {\bibfield  {journal}
  {\bibinfo  {journal} {Rev. Mod. Phys.}\ }\textbf {\bibinfo {volume} {66}},\
  \bibinfo {pages} {1} (\bibinfo {year} {1994})}\BibitemShut {NoStop}%
\bibitem [{\citenamefont {Tserkovnyak}, \citenamefont {Hankiewicz},\ and\
  \citenamefont {Vignale}(2009)}]{Tserkovnyak2009}%
  \BibitemOpen
  \bibfield  {author} {\bibinfo {author} {\bibfnamefont {Y.}~\bibnamefont
  {Tserkovnyak}}, \bibinfo {author} {\bibfnamefont {E.~M.}\ \bibnamefont
  {Hankiewicz}}, \ and\ \bibinfo {author} {\bibfnamefont {G.}~\bibnamefont
  {Vignale}},\ }\href@noop {} {\bibfield  {journal} {\bibinfo  {journal} {Phys.
  Rev. B}\ }\textbf {\bibinfo {volume} {79}},\ \bibinfo {pages} {094415}
  (\bibinfo {year} {2009})}\BibitemShut {NoStop}%
\bibitem [{\citenamefont {Mattis}(2006)}]{Mattis2006}%
  \BibitemOpen
  \bibfield  {author} {\bibinfo {author} {\bibfnamefont {D.~C.}\ \bibnamefont
  {Mattis}},\ }\href@noop {} {\emph {\bibinfo {title} {The theory of magnetism
  made simple}}}\ (\bibinfo  {publisher} {World Scientific publishing},\
  \bibinfo {address} {Hackensack, NJ},\ \bibinfo {year} {2006})\ pp.\ \bibinfo
  {pages} {82--103}\BibitemShut {NoStop}%
\bibitem [{\citenamefont {Nembach}\ \emph {et~al.}(2015)\citenamefont
  {Nembach}, \citenamefont {Shaw}, \citenamefont {Weller}, \citenamefont
  {Ju\'{e}},\ and\ \citenamefont {Silva}}]{Nembach2015}%
  \BibitemOpen
  \bibfield  {author} {\bibinfo {author} {\bibfnamefont {H.~T.}\ \bibnamefont
  {Nembach}}, \bibinfo {author} {\bibfnamefont {J.~M.}\ \bibnamefont {Shaw}},
  \bibinfo {author} {\bibfnamefont {M.}~\bibnamefont {Weller}}, \bibinfo
  {author} {\bibfnamefont {E.}~\bibnamefont {Ju\'{e}}}, \ and\ \bibinfo
  {author} {\bibfnamefont {T.~J.}\ \bibnamefont {Silva}},\ }\href@noop {}
  {\bibfield  {journal} {\bibinfo  {journal} {Nature Physics}\ }\textbf
  {\bibinfo {volume} {11}},\ \bibinfo {pages} {825} (\bibinfo {year}
  {2015})}\BibitemShut {NoStop}%
\bibitem [{\citenamefont {Landau}(1941)}]{landau_theory_1941}%
  \BibitemOpen
  \bibfield  {author} {\bibinfo {author} {\bibfnamefont {L.~D.}\ \bibnamefont
  {Landau}},\ }\href@noop {} {\bibfield  {journal} {\bibinfo  {journal} {J.
  Phys. USSR}\ }\textbf {\bibinfo {volume} {5}},\ \bibinfo {pages} {71}
  (\bibinfo {year} {1941})}\BibitemShut {NoStop}%
\bibitem [{\citenamefont {Whitham}(1974)}]{Whitham1974}%
  \BibitemOpen
  \bibfield  {author} {\bibinfo {author} {\bibfnamefont {G.~B.}\ \bibnamefont
  {Whitham}},\ }\href@noop {} {\emph {\bibinfo {title} {Linear and nonlinear
  waves}}}\ (\bibinfo  {publisher} {John Wiley \& Sons Inc},\ \bibinfo {year}
  {1974})\BibitemShut {NoStop}%
\bibitem [{\citenamefont {Zakharov}\ and\ \citenamefont
  {Ostrovsky}(2009)}]{zakharov_modulation_2009}%
  \BibitemOpen
  \bibfield  {author} {\bibinfo {author} {\bibfnamefont {V.}~\bibnamefont
  {Zakharov}}\ and\ \bibinfo {author} {\bibfnamefont {L.}~\bibnamefont
  {Ostrovsky}},\ }\href@noop {} {\bibfield  {journal} {\bibinfo  {journal}
  {Physica D}\ }\textbf {\bibinfo {volume} {238}},\ \bibinfo {pages} {540}
  (\bibinfo {year} {2009})}\BibitemShut {NoStop}%
\bibitem [{\citenamefont {Frisch}, \citenamefont {Pomeau},\ and\ \citenamefont
  {Rica}(1992)}]{frisch_transition_1992}%
  \BibitemOpen
  \bibfield  {author} {\bibinfo {author} {\bibfnamefont {T.}~\bibnamefont
  {Frisch}}, \bibinfo {author} {\bibfnamefont {Y.}~\bibnamefont {Pomeau}}, \
  and\ \bibinfo {author} {\bibfnamefont {S.}~\bibnamefont {Rica}},\ }\href@noop
  {} {\bibfield  {journal} {\bibinfo  {journal} {Physical Review Letters}\
  }\textbf {\bibinfo {volume} {69}},\ \bibinfo {pages} {1644} (\bibinfo {year}
  {1992})}\BibitemShut {NoStop}%
\bibitem [{\citenamefont {Hoefer}\ and\ \citenamefont
  {Sommacal}(2012)}]{Hoefer2012b}%
  \BibitemOpen
  \bibfield  {author} {\bibinfo {author} {\bibfnamefont {M.}~\bibnamefont
  {Hoefer}}\ and\ \bibinfo {author} {\bibfnamefont {M.}~\bibnamefont
  {Sommacal}},\ }\href {\doibase 10.1016/j.physd.2012.02.003} {\bibfield
  {journal} {\bibinfo  {journal} {Physica D: Nonlinear Phenomena}\ }\textbf
  {\bibinfo {volume} {241}},\ \bibinfo {pages} {890 } (\bibinfo {year}
  {2012})}\BibitemShut {NoStop}%
\bibitem [{\citenamefont {Yan}\ \emph {et~al.}(2013)\citenamefont {Yan},
  \citenamefont {K\'akay}, \citenamefont {Andreas},\ and\ \citenamefont
  {Hertel}}]{Yan2013}%
  \BibitemOpen
  \bibfield  {author} {\bibinfo {author} {\bibfnamefont {M.}~\bibnamefont
  {Yan}}, \bibinfo {author} {\bibfnamefont {A.}~\bibnamefont {K\'akay}},
  \bibinfo {author} {\bibfnamefont {C.}~\bibnamefont {Andreas}}, \ and\
  \bibinfo {author} {\bibfnamefont {R.}~\bibnamefont {Hertel}},\ }\href@noop {}
  {\bibfield  {journal} {\bibinfo  {journal} {Phys. Rev. B}\ }\textbf {\bibinfo
  {volume} {88}},\ \bibinfo {pages} {220412} (\bibinfo {year}
  {2013})}\BibitemShut {NoStop}%
\bibitem [{\citenamefont {Sparks}(1965)}]{Sparks1965}%
  \BibitemOpen
  \bibfield  {author} {\bibinfo {author} {\bibfnamefont {M.}~\bibnamefont
  {Sparks}},\ }\href@noop {} {\emph {\bibinfo {title} {Ferromagnetic relaxation
  theory}}}\ (\bibinfo  {publisher} {McGraw-Hill},\ \bibinfo {year}
  {1965})\BibitemShut {NoStop}%
\bibitem [{\citenamefont {Braun}(2012)}]{Braun2012}%
  \BibitemOpen
  \bibfield  {author} {\bibinfo {author} {\bibfnamefont {H.-B.}\ \bibnamefont
  {Braun}},\ }\href {\doibase 10.1080/00018732.2012.663070} {\bibfield
  {journal} {\bibinfo  {journal} {Advances in Physics}\ }\textbf {\bibinfo
  {volume} {61}},\ \bibinfo {pages} {1} (\bibinfo {year} {2012})}\BibitemShut
  {NoStop}%
\end{thebibliography}

\clearpage
\begin{center}\section*{Supplementary material}\end{center}

\section{Nondimensionalization of the Landau-Lifshitz equation}

The Landau-Lifshitz (LL) equation can be written as
\begin{equation}
  \label{eq:SM1}
  \frac{\partial \mathbf{M}}{\partial t} =
  -\frac{\gamma\mu_0}{1+\alpha^2}\left[\mathbf{M}\times\mathbf{H}_\mathrm{eff}+
    \frac{\alpha}{M_s}\mathbf{M}\times
    \left(\mathbf{M}\times\mathbf{H}_\mathrm{eff}\right)\right],  
\end{equation}
where the damping $\alpha$ is in Gilbert form, $\mathbf{M}$ is the magnetization vector, $M_s=|\mathbf{M}|$ is the saturation magnetization, $\gamma$ is the gyromagnetic ratio, $\mu_0$ is the vacuum permeability, and the effective field is
\begin{equation}
\label{eq:SM2}
  \mathbf{H}_\mathrm{eff} = \lambda_\mathrm{ex}^2M_s\Delta\mathbf{m}+M_\mathrm{eff}(\mathbf{m}\cdot\hat{\mathbf{z}})\hat{\mathbf{z}}+H_0\hat{\mathbf{z}},
\end{equation}
where $\mathbf{m}=\mathbf{M}/M_s$, $\lambda_\mathrm{ex}=\sqrt{2A/(\mu_0M_s^2)}$ is the exchange length, $A$ is the exchange constant with units Jm$^{-1}$, ${M_\mathrm{eff}=(2K_u/(\mu_0M_s)-M_s)}$, and $K_u$ is the perpendicular magnetic anisotropy energy density with units Jm$^{-3}$. In order to generalize our results for materials favoring either planar or uniaxial anisotropy, we introduce the quantity $\sigma=\mathrm{sgn}(M_s-H_k)$, where $H_k=2K_u/(\mu_0M_s)$ is the perpendicular magnetic anisotropy field. It is then possible to rescale time ${t\rightarrow\gamma\mu_0|H_k-M_s|(1+\alpha^2)^{-1} t}$, space ${x\rightarrow\sqrt{|H_k/M_s-1|}\lambda_{ex}^{-1}x}$, and field $h_\mathrm{eff}=H_\mathrm{eff}/|H_k-M_s|$. As a result, we obtain the dimensionless Landau-Lifshitz equation
\begin{equation}
\label{eq:SM3}
\frac{\partial \mathbf{m}}{\partial t} =
-\mathbf{m}\times\mathbf{h}_\mathrm{eff}-\alpha\mathbf{m}\times\mathbf{m}
\times\mathbf{h}_\mathrm{eff}, 
\end{equation}
[Eq.~(1) in the main text]. Note that the special case of an isotropic ferromagnet $H_k=M_s$ leads to a divergence from our nondimensionalization. However, we consider that an isotropic thin film ferromagnet is experimentally unlikely, in which case the divergence does not affect the generality of the analytical results.

\section{Derivation of hydrodynamic equations}

The dispersive hydrodynamic representation can be derived by inserting the canonical Hamiltonian transformation
\begin{subequations}
\label{eq:SM4}
\begin{eqnarray}
\label{eq:SM41}
n &=& m_z,\\
\label{eq:SM42}
\mathbf{u} &=& - \nabla \Phi = -\nabla \left [
  \arctan{\left(m_y/m_x\right)} \right ],
\end{eqnarray}
\end{subequations}
into Eq.~\eqref{eq:SM3}. Solving for $\partial\Phi/\partial t$ we obtain the exact transformation
\begin{equation}
  \label{eq:SM5}
    \frac{\partial \Phi}{\partial t} = -(\sigma - | \mathbf{u} |^2) n  + \frac{\Delta n}{1 - n^2} + \frac{ n |\nabla n|^2}{(1-n^2)^2}
    +    h_0- \frac{\alpha}{1 - n^2} \nabla \cdot [(1-n^2) \mathbf{u}].
\end{equation}

Solving for $\partial n/\partial t$ and $\partial\mathbf{u}/\partial t=-\nabla\partial\Phi/\partial t$ we obtain
\begin{subequations}
\label{eq:SM6}
\begin{eqnarray}
  \label{eq:SM61}
    \frac{\partial n}{\partial t} &=&
    \nabla\cdot\left[(1-n^2)\mathbf{u}\right] +
    \alpha(1-n^2)\frac{\partial \Phi}{\partial
        t} ,\\
  \label{eq:SM62}
  \frac{\partial \mathbf{u}}{\partial t} &=&
  \nabla\left[(\sigma-|\mathbf{u}|^2)n
    \right] 
  - \nabla\left[\frac{\Delta n}{1-n^2}+\frac{n |\nabla
        n|^2}{(1-n^2)^2}\right]\nonumber\\
				&&-\nabla h_0
  +\alpha\nabla\left[\frac{1}{1-n^2}\nabla\cdot\left[(1-n^2) 
        \mathbf{u}\right]\right].
\end{eqnarray}
\end{subequations}

Interpreting $n$ as the longitudinal spin density, we observe that Eq.~\ref{eq:SM61} is reminiscent of the continuity equation~\cite{Zhang2005} in the absence of damping
\begin{equation}
\label{eq:SMZhang}
  \frac{\partial \mathbf{M}\cdot\hat{\mathbf{z}}}{\partial t} = -\nabla\cdot\mathbf{J}_s + T + T_\mathrm{relax},
\end{equation}
where $\mathbf{J}_s$ is the spin current, $T$ is spin torque, and $T_\mathrm{relax}$ is a relaxation process. Therefore, we identify the first term of the right-hand-side of Eq.~\eqref{eq:SM61} as the spin current, acknowledging the sign difference arising from our definition of $\mathbf{u}$.

The term $(1-n^2)$ appearing in Eq.~\eqref{eq:SM61} is non-negative, quantifying the fluid density deviation from
vacuum, $|n|=1$. However, one can verify that this is
not a conserved quantity. In fact, neglecting damping,
\begin{equation}
  \frac{\partial(1-n^2)}{\partial t}=2n\frac{\partial n}{\partial t} =
    2n\nabla\cdot\left[(1-n^2)\mathbf{u}\right],
\end{equation}
whose right-hand-side cannot be written as a divergence. This agrees
with the notion that spin current itself is not a conserved quantity even when damping is zero, whereas
the longitudinal spin density is always conserved~\cite{Zhang2005} in the absence of specific processes. Indeed, this is a restatement of the fact that spin currents in ferromagnets have both a coherent and diffusive character, where the coherent form is typified by spin wave propagation, whereas the diffusive character is associated with thermal magnon kinetics. As such, the lack of conservation for the fluid density in ferromagnets is analogous to the fact that the energy-conserving coherence lifetime $\tau_2$ in the Bloch formulation of spin dynamics can be substantially shorter than the thermodynamic relaxation time $\tau_1$~\cite{Sparks1965}.

In the long-wavelength, low-frequency, near saturation density limit, $|\nabla n| \ll 1$, $|\mathbf{u}|^2 \ll 1$, and
$n\ll1$, we obtain
\begin{subequations}
\label{eq:SM7}
\begin{eqnarray}
  \frac{\partial n}{\partial t} &=& -\Delta\Phi+\alpha\frac{\partial\Phi}{\partial t},\\
	\frac{\partial\Phi}{\partial t} &=& -\sigma n+h_0+\alpha\Delta\Phi.
\end{eqnarray}
\end{subequations}
These equations are consistent with the equations describing spin superfluidity~\cite{Sonin2010,Takei2014}.

Considering a UHS with a uniform precessional frequency
$d\Phi/d t=\Omega$, no damping, and neglecting spatial
variation in $n$, the magnetic analogue of Bernoulli's equation can be
obtained by multiplying Eq.~\eqref{eq:SM5} by $n/2$
\begin{equation}
  \label{eq:SM8}
  P(n,|\mathbf{u}|) + \frac{1}{2} |\mathbf{u}|^2 + \frac{1}{2} n(\Omega - h_0) =
  -\frac{\sigma}{2}, 
\end{equation}
where $P$ is pressure, the second term is the kinetic energy, the third term is a potential (analogous to gravity).

\section{Micromagnetic simulations}

We perform micromagnetic simulations to validate our analytical
results. For the magnetic defect in the static frame, $\mathbf{V}=0$,
we use the GPU-accelerated code Mumax3~\cite{Vansteenkiste2014}. In
order to maintain a SDW in the absence of external forcing, e.g., spin
injection, we set a SDW with an integer number of periods as an
initial magnetization state and implement periodic boundary conditions
(PBC) along the nanowire's length and free spin boundary
  conditions transversely. The integer period
requirement sets the length of the simulation area. We use idealized
physical parameters consistent with infinitely permeable Permalloy,
$M_s = 790$~kA/m, $A=10$~pJ/m, and $H_k=0$. In this case, space is
normalized by $\lambda_\mathrm{ex}\approx5$~nm according to the
  nondimensionalization introduced above. Three types of simulations
were performed:
\begin{enumerate}
\item To estimate the sonic curve, we set a nanowire of
  $20$ in width, $20\pi/\bar{u}$
  in length, and $1$ in thickness. A SDW at a
  particular $\bar{u}$ is defined as an initial condition. Taking
  advantage of the field relaxation of the longitudinal spin density, we set $\alpha=0.01$ and sweep the field at a
  normalized rate of $1\times10^{-5}$ (on the order of
  $0.3$~T/$\mu$s). This rate is slow enough to sweep $\bar{n}$ as a
  function of field. A point-defect is introduced as an absence of
  magnetization. The supersonic condition is obtained at the point
  when the linear increase in $\bar{n}$ changes slope, corresponding
  to a quantized drop of the SDW topological charge.
\item The supersonic transition is also estimated by including
  non-local dipolar fields. In contrast to case 1 above, we do not
  consider any defect and instead add room temperature ($300$~K)
  thermal fluctuations. The supersonic transition is identified as a
  dramatic drop in the longitudinal spin density $n = m_z$,
  corresponding to the generation of spin wave as the UHS decays. To take into account the noise introduced by thermal
  fluctuations, a linear fit is applied to the averaged $m_z$. The
  error is obtained from the standard deviation throughout the
  simulation time.
\item The effect of finite-sized obstacles is obtained from
  simulations on a domain spanning $640$ in width,
  $640$ in length, and $1$ in
  thickness. We set the damping to $\alpha=0.01$ and utilize an
  external field to maintain the desired longitudinal spin
  density. This size guarantees that the generated features damp
  before reaching the simulation physical boundary.
\end{enumerate}

In the case of the moving obstacle, a localized applied field spot, we
use a pseudo-spectral method to solve the two-dimensional
Landau-Lifshitz equation in the moving frame~\cite{Hoefer2012b}. The
simulation is initialized with an homogeneous state,
$\mathbf{u}=0$. The localized field spot magnitude is set to $0.05$ in
normalized units. Two types of simulations are performed:
\begin{enumerate}
\item To estimate the sonic curve, we set a simulation area of $100$
  in width and $80$ in length. A point-field is implemented as a
  hyper-Gaussian of order 4 and width $1$. Transient structures are
  damped during the simulation time. Thus, the supersonic transition
  is estimated by the existence of a well-defined wave front, i.e.,
  wavefront radiation.
\item The effect of finite-sized obstacles is obtained from a
  simulation following the geometry described in point 3 above.
\end{enumerate}

\section{Fluid-like behavior for the in-plane magnetization}

\begin{figure}
  \centering \includegraphics[width=5.5in]{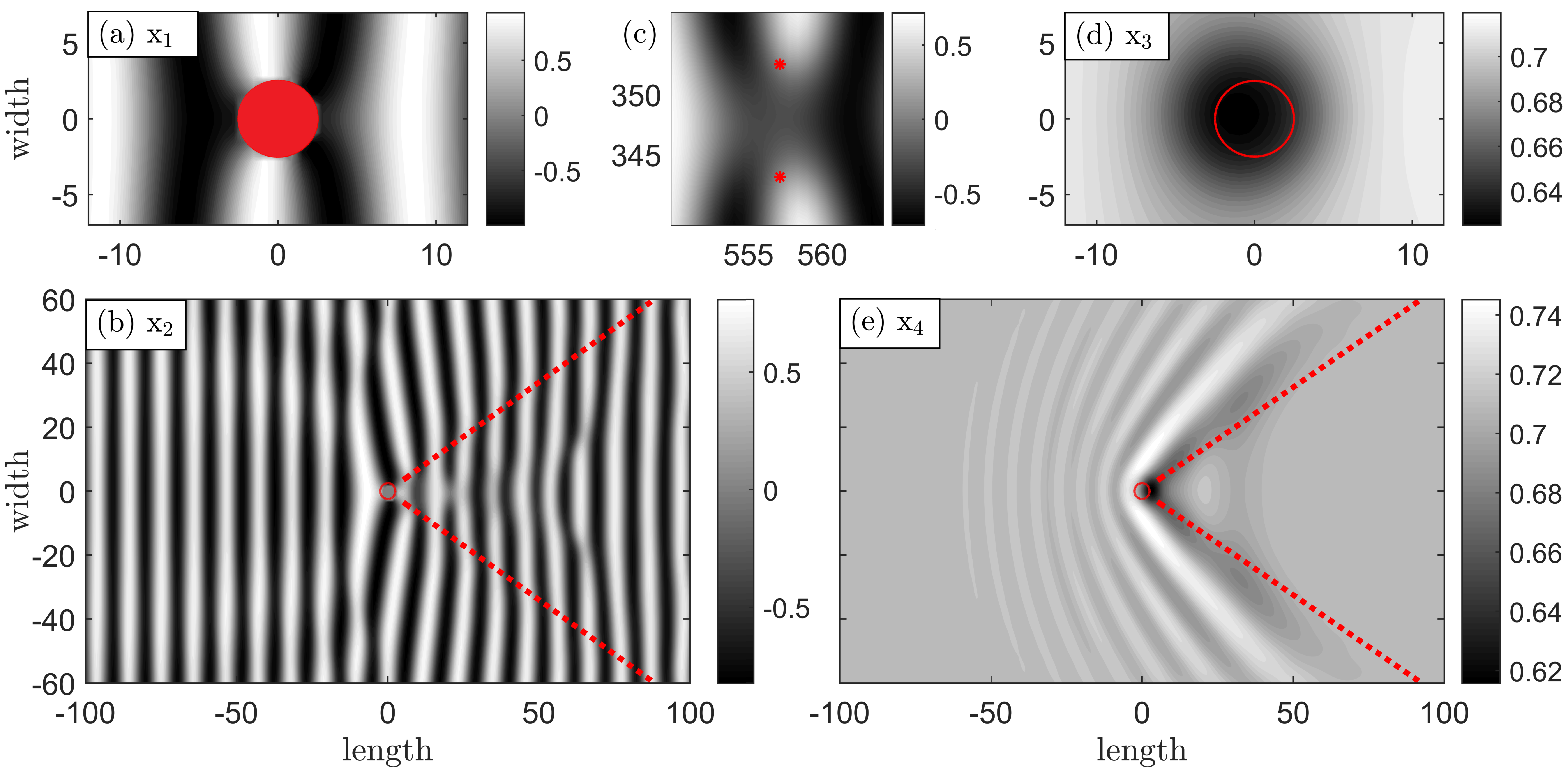}
  \caption{ \label{fig3Supp} Snapshots of (a), (b) a SDW flowing past a stationary magnetic defect
    ($\bar{V}=0$) and; (d), (e) a homogeneous state subject to a moving, localized magnetic
    field ($\bar{V}\neq0$). The grayscale map shows the $m_x$ component and the panels are in exact correspondence with those shown in the main text. }
\end{figure}

In the main text, Fig.~3 shows the fluid-like behavior of a thin film planar ferromagnet, both in the static and moving reference frames. A hydrodynamic visualization is chosen, where the longitudinal spin density $n$ and fluid velocity $\mathbf{u}$ are represented by a grayscale map and vector field, respectively. Here, we show the corresponding $\hat{\mathbf{x}}$ magnetization component ($m_x$) in Fig.~\ref{fig3Supp} as a grayscale map. Note that the SDW, Fig.~\ref{fig3Supp}(a-c), is visualized as a periodic variation of $m_x$. The features arising in the subsonic and supersonic conditions distort this periodic configuration. Notably, the vortex-antivortex, Fig.~\ref{fig3Supp} breaks the periodicity of the SDW. This can be understood by topological arguments, as outlined in the next section. In contrast, the homogeneous state in the moving reference frame exhibits similar features for $m_x$ and $n$.

\section{Hydrodynamic topology}

The notion of topology in our hydrodynamic framework is tightly linked
to the definition of the magnetic ground state. The concept of
topology is typically used to mathematically classify states or
textures into groups separated by an infinite barrier (see
  Ref.~\onlinecite{Braun2012} for details). Thus, a texture
belonging to a certain topological group cannot change its topological
group i.e., it is topologically protected. Within a topological
group, textures are allowed to be continuously deformed, i.e., such a
deformation does not cross a barrier.

\begin{figure}[t]
\centering \includegraphics[width=6in]{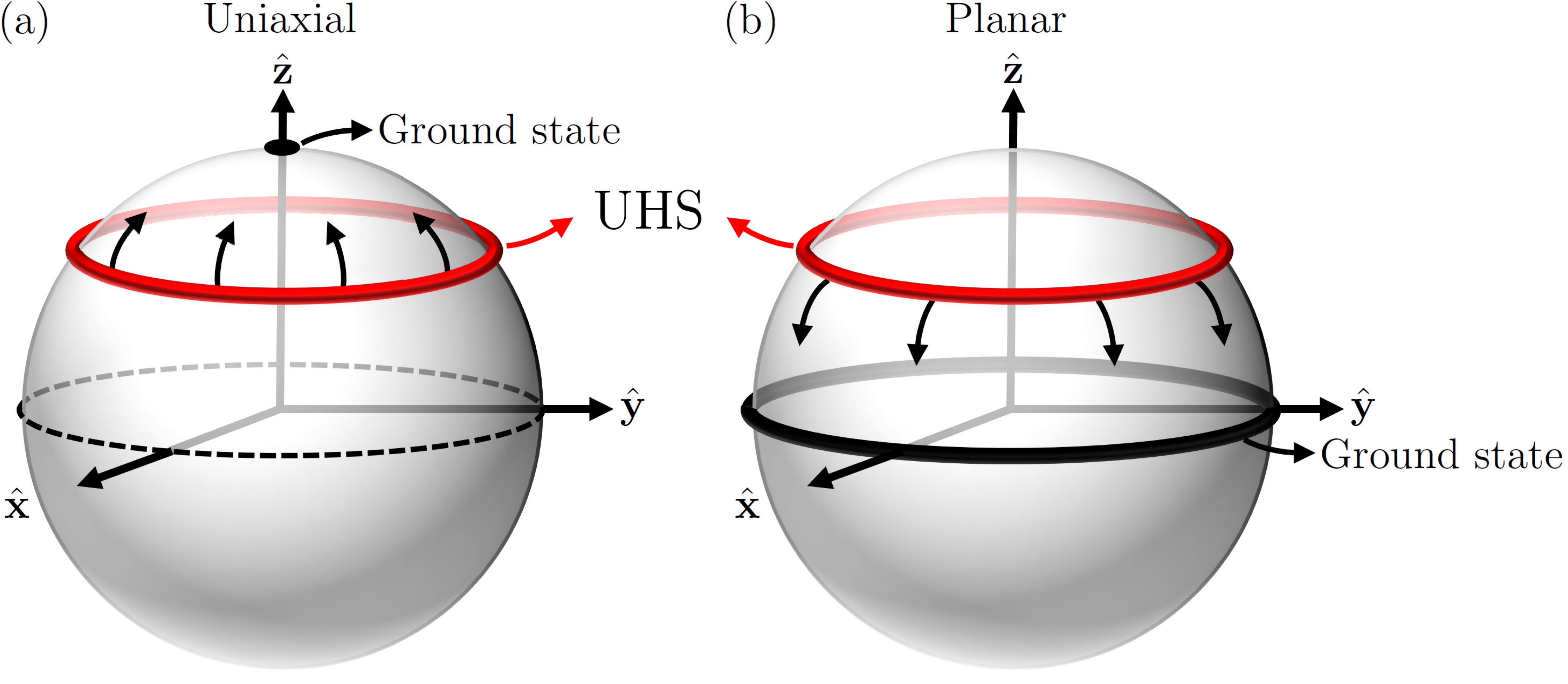}
\caption{ \label{figSM2} Relaxation of a UHS in a (a) uniaxial and (b) planar ferromagnet to their ground state. In (a), the UHS can relax to the homogeneous magnetization state whereas in (b) the UHS maintains its phase structure and, therefore, its topology. }
\end{figure}
Physically, topological barriers correspond to a large, but finite,
energy. In a magnetic system, such a barrier can be interpreted as the
magnetization hard axis. For example, a domain wall separating two
magnetic states crosses a hard axis, and is therefore topologically
protected. In fact, domain walls are usually annihilated by hitting a
pinning site or a physical boundary, where the energy barrier can be
easily surpassed.

In the hydrodynamic context, we are interested in solutions such as a
UHS, exhibiting phase rotations in the
$\hat{\mathbf{x}}$-$\hat{\mathbf{y}}$ plane. Therefore, in the
magnetization unit sphere, a UHS traces a circle~\cite{Sonin2010}. At
this point, it is important to distinguish between uniaxial and planar
ferromagnets. For uniaxial ferromagnets, the UHS circle can be
continuously deformed along the unit sphere latitude until it reaches
its ground state, the $\hat{\mathbf{z}}$ axis. This is schematically
shown in Fig.~\ref{figSM2}(a) for a UHS circle (red) in the northern
hemisphere. Consequently, UHSs in uniaxial ferromagnets do not cross
the hard axis to return to the homogeneous ground state,
$\pm\hat{\mathbf{z}}$, which is typically referred to as a state with
trivial topology. It is important to point out that other topological
structures are allowed in uniaxial ferromagnets, such as Bloch domain walls,
where the ground states are separated by the hard plane.

In the case of planar ferromagnets, the ground state is the isotropic
easy plane. Thus, a UHS can continuously deform to a circle along the
equator, i.e., a SDW when $h_0=0$, as shown in
Fig.~\ref{figSM2}(b). Since this ground state is not homogenous, it
possesses a non-trivial topology. For this reason, UHSs in planar
ferromagnets exhibit topological protection. The only mechanism to
annihilate a UHS is to cross the hard axis i.e., the
$\pm\hat{\mathbf{z}}$ poles. As shown in the Supplementary Video 1,
this is achieved by creating a magnetic vortex whose core points
exactly along the hard axis and it is therefore an avenue to unwind a
phase rotation. From this topological argument, it is also clear that
the state with $n=1$ does not possess topology and it is therefore the
hydrodynamic vacuum state.

\end{document}